\documentclass[%
 reprint,
superscriptaddress,
 amsmath,amssymb,
 aps,
prb,
]{revtex4-2}
\usepackage{chemformula}
\usepackage{graphicx}
\usepackage{dcolumn}
\usepackage{bm}

\def \mc{\mathcal}

\begin{document}

\preprint{APS/123-QED}

\title{Theory of superconductivity mediated by Rashba coupling in incipient ferroelectrics}

\author{Maria N. Gastiasoro}
\email{maria.ngastiasoro@uniroma1.it}
\affiliation{ISC-CNR and Department of Physics, Sapienza University of Rome, Piazzale Aldo Moro 2, 00185, Rome, Italy}
\author{Maria Eleonora Temperini}%
\affiliation{ISC-CNR Institute for Complex Systems, via dei Taurini 19, 00185 Rome, Italy}%
\author{Paolo Barone}
\affiliation{SPIN-CNR Institute for Superconducting and other Innovative Materials and Devices, Area della Ricerca di Tor Vergata, Via del Fosso del Cavaliere 100, 00133 Rome, Italy}
\author{Jose Lorenzana}
\affiliation{ISC-CNR and Department of Physics, Sapienza University of Rome, Piazzale Aldo Moro 2, 00185, Rome, Italy}%

\date{\today}

\begin{abstract}
  Experimental evidence suggests that superconductivity in \ch{SrTiO3} is mediated by a soft transverse ferroelectric mode which, according to conventional theories, has negligible coupling with electrons. A phenomenological Rashba type coupling has been proposed on symmetry arguments but a microscopic derivation is lacking.  Here we
  fill this gap and obtain a linear coupling directly from a minimal 
  microscopic 
  model of the electronic structure. We find that the effective electron-electron pairing interaction has a strong momentum dependence. This yields an unusual situation in  which the leading $s$-wave channel is followed by a sub-leading $p$-wave state which shows a stronger pairing instability than the $d$-wave state.
  The bare Rashba coupling constant is estimated for the lowest band of doped SrTiO$_3$ with the aid of first-principles computations. Extrapolating the estimation to the multi-band regime we show that it can explain bulk superconductivity. For densities where the Meissner effect has not been observed 
  an extra softening due to inhomogeneities is needed to explain the zero resistance state.
\end{abstract}

\maketitle



The role of polar fluctuations in metallic systems close to a transition where a polar axis develops has recently attracted considerable attention.
Across this structural transition, inversion symmetry is broken through polar displacements of atoms and the system becomes ferroelectric (FE) if there are no free carriers, or a polar metal if the transition happens in the presence of a Fermi sea. In both cases, the development of the polar axis can often be described by the softening of an infrared-active phonon associated to the relevant polar displacements. The coupling of the electronic Fermi sea to this soft mode and its feedback has been central to many recent theoretical and experimental works. These include superconductivity mediated by FE fluctuations~\cite{Edge2015,Kozii2019,van2019possible,van2019possible,Gastiasoro2020,hameed2020ferroelectric,yoon2021low}, the possibility of attraction in the odd-parity Cooper channel~\cite{Fu2015,Martin2017,schumann2020possible,Sumita2020}, unusual transport properties~\cite{zhou2018electron,wang2019charge,kumar2021,collignon2021quasi,yue2021}, 
and novel quantum critical phenomena~\cite{Volkov2020}.

SrTiO$_3$ (STO) is a model incipient ferroelectric where polar quantum fluctuations are believed to be strong. It becomes superconducting (SC) upon carrier doping, and experimentally it has been widely reported that tuning the system even closer to the polar transition notably enhances the SC critical temperature $T_c$~\cite{Stucky2016,Rischau2017,Herrera2019,Tomioka2019,Ahadi2019,Harter2019,Enderlein2020,franklin2021giant,salmani2021}. This has led to a substantial amount of theoretical works to consider the exchange of polar fluctuations as a potential candidate for the Cooper pairing mechanism ~\cite{Edge2015,wolfle2018,wolfle2019reply,Arce-Gamboa2018quantum,kedem2018novel,van2019possible,Gastiasoro2020,Sumita2020,Gastiasoro2020review,volkov2021superconductivity,kiseliov2021theory}.

There are two important issues that remain to be addressed regarding this novel pairing mechanism. The first is the influence of the dynamics of the polar fluctuations on the superconducting instability, known to be important in the well studied cases of antiferromagnetic and ferromagnetic critical fluctuations. The second challenge, which we address in this paper, is providing a mechanism that allows for a significant coupling between the soft polar mode and the electronic degrees of freedom. 
Because the long-range dipolar interaction pushes the longitudinal optical mode to higher energies in the long-wavelength limit, the soft mode has a transverse polarization, which implies no gradient-like linear coupling, such as Fr\"ohlich coupling, the standard electron-phonon interaction considered in polar crystals. Even including longitudinal components from crystal field corrections to the polarization of the soft mode leads to a negligible coupling~\cite{ruhman2019comment}. Different alternative mechanisms have also been proposed, including a revival of an old idea involving the quadratic coupling to the soft mode~\cite{Ngai,van2019possible,volkov2021superconductivity,kiseliov2021theory} as well as the dynamical screening of the Coulomb interaction due to longitudinal modes~\cite{Takada1980,Ruhman2016,Enderlein2020}.

Here we focus on an alternative linear coupling mechanism to the FE mode which was proposed using symmetry arguments in the presence of spin-orbit coupling (SOC)~\cite{Fu2015,Kanasugi2019,Kozii2019} or in multi-band systems~\cite{Volkov2020}. These theories are semi-phenomenological in that they do not completely determine the momentum dependence of the coupling nor its magnitude.
We present a minimal microscopic model for the linear coupling to a polar transverse mode in a system with SOC and derive the corresponding Rashba-like electron-phonon coupling Hamiltonian. We show that unlike what is found in models of Rashba splitting at surfaces~\cite{Petersen2000,Khalsa2013,Zhong2013}, the interaction does not scale with the atomic spin-orbit parameter $\lambda_{SO}$, but rather with a conventional electron-phonon matrix element.
With the aid of density functional theory (DFT) frozen-phonon computations we estimate the bare Rashba-like coupling for doped STO and find it to be substantial. Projecting the interaction into the Cooper channel, we study the pairing  weak-coupling solutions for both even and odd-parity channels. Using the \emph{ab initio} results for the bare Rashba coupling allows us to obtain an estimate for the BCS coupling $\lambda$ of the leading singlet channel in STO. The coupling explains bulk superconductivity observed at relatively high density. In the very dilute regime, where the Meissner effect is absent, it can provide the right order of magnitude of the pairing temperature in the presence of inhomogeneities.

\begin{figure}
    \centering
    \includegraphics[width=\linewidth]{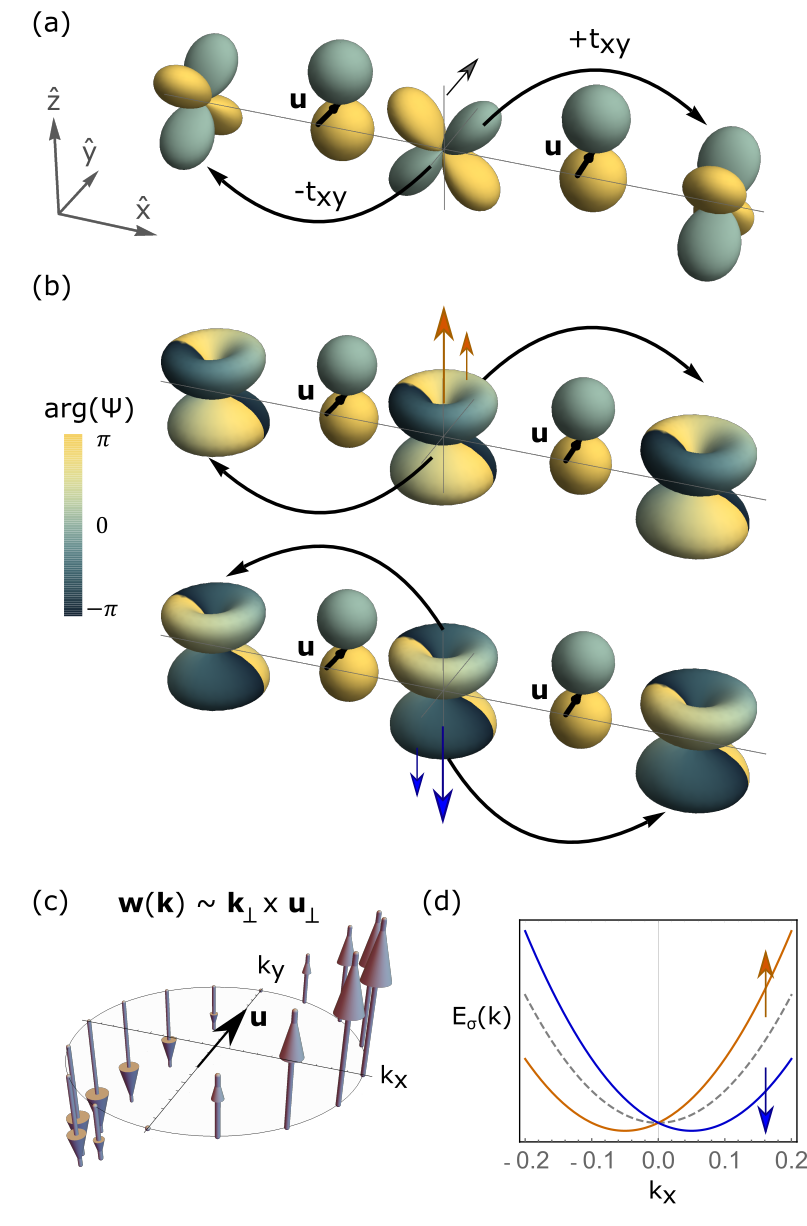}
    \caption{(a) Induced inter-orbital hopping $t_{xy}$ [Eq.~\eqref{eq:tmunukq0}] between $3d$ orbitals with spin-$1/2$ (gray arrow) mediated by $p_z$ orbitals. It changes sign along the bond $\bm{\hat{x}}$, with a finite polar displacement along $\bm{\hat{y}}$, $u^{\mathrm{O}_x}_y-u^{\mathrm{TM}}_y\neq 0$. 
    (b) Same process in the SOC basis $|\frac{3}{2},\pm \frac{3}{2} \rangle$  [Eq.~\eqref{eq:soc}] with spin-dependent $c^\dagger_\uparrow c_\uparrow$ (top) and $c^\dagger_\downarrow c_\downarrow$ (bottom). The orange and blue arrows show the polarization of the spin along $\bm{\hat{z}}$, parallel to the angular momentum. The color legend represents the argument of the wave function. 
    (c) Rashba spin-orbit field $\bm{w}(\bm k)$ (see text) and (d) SOC band spin-splitting $ E_\sigma(\bm{k})$, in the presence of the polar displacement shown in (a). The gray dashed line represents the doubly-degenerate SOC bands when $\bm{u}=0$. }
    \label{fig:model}
\end{figure}

\emph{Minimal model.} We consider doped oxide perovskite compounds in the vicinity of a FE instability, where the soft polar mode may play a prominent role. 
The conduction band is formed from $t_{2g}$ orbitals of the transition-metal (TM) atom $d_{yz}$, $d_{zx}$, and $d_{xy}$, which we denote respectively $\mu=x,y,$ and $z$. The inclusion of atomic SOC, which is crucial for the Rashba coupling~\cite{Fu2015,Kozii2015,Kanasugi2018,Kozii2019,Kanasugi2019} to the soft FE phonon derived below, mixes the $t_{2g}$ orbitals into SOC bands $|j,j_z\rangle$ with total angular momentum $j=\frac{3}{2}$ and $j=\frac{1}{2}$ 
\cite{Bistritzer2011,Zhong2013}.
A polar distortion of the lattice introduces new hopping channels~\cite{Khalsa2013,Djani2019,Volkov2020}  
\begin{equation}
    \mc{H}_u= \sum_{\bm{k} \bm{q} \mu \nu \sigma\sigma'} t^{\sigma\sigma'}_{\mu\nu}(\bm{k},\bm{q}) c^\dagger_{\mu\sigma}(\bm{k}+\frac{\bm{q}}{2}) c_{\nu\sigma'}(\bm{k}-\frac{\bm{q}}{2})+\mathrm{h.c.}
    \label{eq:Hu}
\end{equation}
between a d-orbital $\mu$ with spin $\sigma$ and a nearest neighbor d-orbital $\nu$ with spin $\sigma'$, mediated by the p-orbitals of the bridging O atom. These induced hopping elements are forbidden in the distortion-free system. Following Slater-Koster rules~\cite{slater} we assume the induced hopping depends only on the displacements of the two TM atoms $\bm{u}^{\mathrm{TM}}$ involved in the process and the O atom $\bm{u}^{\mathrm{O}_\delta}$ on the TM-TM bond $\bm{\delta}$, as illustrated in Fig.~\ref{fig:model}(a). 
In particular, spin conserving processes with $\sigma'=\sigma$ result in 
a simple expression for the inter-orbital ($\mu\neq\nu$) hopping to linear order in the displacement.
In the $\bm{k},\bm{q}\rightarrow 0$ limit the induced hopping becomes (see Appendix~\ref{app:inter})
\begin{eqnarray}
     t_{\mu\nu}(\bm{k},\bm{q})\approx&& 2it'_{\mu\nu} \left( \left(\bm{\hat\mu}\times\bm{\hat\nu}\right)\times\bm k\right)\cdot\left(\bm{u}^{\mathrm{O}}(\bm{q})- \bm{u}^{\mathrm{TM}}(\bm{q})\right)
    \nonumber \\
     &&+ it'_{\mu\nu} \left(q_{\mu} u_{\nu}^{\mathrm{TM}}(\bm{q})+q_{\nu} u_{\mu}^{\mathrm{TM}}(\bm{q})\right) +\mc{O}(k^3)
     \label{eq:tmunukq0}
\end{eqnarray}
Here, $\bm{\hat{\mu}},\bm{\hat{\nu}}=\bm{\hat{x}},\bm{\hat{y}},\bm{\hat{z}}$ is a unitary vector normal to the plane spanned by the $t_{2g}$ orbital $\mu, \nu$, and we used the simplified notation $t'_{\mu\nu}=\frac{\partial t_{\mu\nu}}{\partial u^{\mathrm{O}_\mu}_\nu}=-\frac{\partial t_{\mu\nu}}{\partial u^{\mathrm{TM}}_\nu }$ .
The first term in Eq.~\eqref{eq:tmunukq0} is proportional to the polarization vector $\bm{P}(\bm{q})=Z\left[\bm{u}^{\mathrm{O}}(\bm{q})- \bm{u}^{\mathrm{TM}}(\bm{q})\right]\equiv Z \bm{u}(\bm{q})$ with an effective charge $Z$ per unit cell volume, and thus describes inter-orbital coupling induced by polarization waves.
Figure~\ref{fig:model}(a) illustrates the induced hopping between orbitals $\mu=x$ and $\nu=y$ for a finite polar displacement along $\bm{\hat{y}}$, which changes sign with hopping direction. As we explicitly show below, this process is responsible for the Rashba coupling. The second term in Eq.~\eqref{eq:tmunukq0}, instead, corresponds to the coupling to strain gradients as produced by acoustic modes and vanishes at the zone center. 

In order to keep the interaction with the lattice as simple as possible, we focus on the $|\frac{3}{2},\pm \frac{3}{2} \rangle$ SOC band, 
\begin{subequations}
 \label{eq:soc}
\begin{eqnarray}
     &c_{+\frac{3}{2}}=(c_{x\uparrow}+ i c_{y\uparrow})/\sqrt{2} \equiv c_\uparrow,\\
     &c_{-\frac{3}{2}}=(c_{x\downarrow}- i c_{y\downarrow})/\sqrt{2}\equiv c_\downarrow,
\end{eqnarray}
\end{subequations}
formed by degenerate states $j_z=\pm\frac{3}{2}$ with opposite angular momentum and spin [see Fig.~\ref{fig:model}(b)]
and involving only two of the three $t_{2g}$ orbitals. This situation describes the first available electron doped state in STO below $105$ K, where a tetragonal distortion splits the $t_{2g}$ levels (see Appendix~\ref{app:dft}). The derivation of the coupling to other SOC bands $|j,j_z\rangle$ is analogous. 

Projecting Eqs.~\eqref{eq:Hu} and \eqref{eq:tmunukq0} to the spinor of the 
SOC band Eq.~\eqref{eq:soc}, $\psi^{\dagger}=(c^{\dagger}_{\uparrow},c^{\dagger}_{\downarrow})$, we arrive at the deformation potential expression $\mc{H}_u= \sum_{\bm{k} \bm{q} }  \psi^\dagger(\bm{k}+\frac{\bm{q}}{2}) \bm{\Lambda}(\bm{k},\bm{q}) \psi(\bm{k}-\frac{\bm{q}}{2})$ with  
\begin{equation}
    \bm{\Lambda}(\bm{k},\bm{q})=2 t'_{xy} \bm{\hat{z}}\times\bm{k} \sigma_z\cdot \left[\bm{u}^{\mathrm{O}}(\bm{q})- \bm{u}^{\mathrm{TM}}(\bm{q})\right].
    \label{eq:Gamma1}
\end{equation}
Here, $\bm{\Lambda}(\bm{k},\bm{q})$ is a 2$\times$2 matrix in pseudospin space, with the Pauli matrix $\sigma_z$. 
Hence, we obtained a coupling between a
polar displacement $\bm{u}$ with the pseudospin of the electrons. Note that only the polarization-wave coupling term
in Eq.~\eqref{eq:tmunukq0}, contributes to this Rashba deformation potential.

Interestingly, the SOC parameter $\lambda_{SO}$ does
not appear in Eq.~(\ref{eq:Gamma1}). This is because near the zone center the angular momentum is not quenched and states with well defined $j_z$ [Eq.~\eqref{eq:soc}]
are split already at zero order in $\lambda_{SO}$ from the rest of the states. Because of their chiral nature, these states propagate with different velocity in the presence of the polar mode 
without further intervention of the SOC as schematized in Figs.~\ref{fig:model} (b) and (d).
In order to better describe this effect, we rewrite the deformation potential matrix Eq.~\eqref{eq:Gamma1} in a more conventional form, i.e., in terms of the Rashba field $\bm{w}(\bm{k})$ experienced by the SOC electrons, that couples to their pseudospin $\bm{\sigma}$: $ \bm{\Lambda}(\bm{k},\bm{q}=0)=2 t'_{xy}u_0\left( \bm{k}_\perp \times \bm{\hat{u}}_{\perp} \right) \cdot \bm{\sigma} \equiv \bm{w}(\bm{k})\cdot \bm\sigma$. 
Here, 
$\bm{k}_\perp$ and $\bm{\hat{u}}_\perp$ are vectors in the $xy$ plane, and 
we have chosen a 
uniform polar displacement $\bm{u}(0)=u_0\bm{\hat{u}}$. The Rashba field $\bm{w}(\bm{k})$ for the polar displacement $\bm{\hat{u}}= \bm{\hat{y}}$, shown in Fig.~\ref{fig:model}(c), results in the characteristic pseudospin band splitting of Fig.~\ref{fig:model}(d). 

Finally, in order to obtain the electron-FE-phonon Hamiltonian we quantize the displacements $\bm{u}(\bm{q})$ in Eq.~\eqref{eq:Gamma1} and decompose them in a set of normal modes $\lambda$:
\begin{eqnarray}
    \label{eq:Helph}
    &&\mc{H}_u= \frac{1}{\sqrt{N}}\sum_{\bm{k} \bm{q} \lambda}  \mathfrak{g}_\lambda(\bm{k},\bm{q}) \psi^\dagger(\bm{k}+\frac{\bm{q}}{2}) \sigma_z \psi(\bm{k}-\frac{\bm{q}}{2}) \mc{A}_{\bm{q}\lambda}
     \\
    &&\mathfrak{g}_\lambda(\bm{k},\bm{q})= 2 t'_{xy}\sqrt{\frac{\hbar}{2 \mu_S \omega_{\bm{q}\lambda}}} \bm{\hat{z}}\times\bm{k}\cdot \bm{\hat{n}}_\lambda(\bm{q}) 
    \label{eq:glambda}
\end{eqnarray}
Here $\mc{A}_{\bm{q}\lambda}=\left(a_{\bm{q}\lambda}+a^{\dagger}_{-\bm{q}\lambda}\right)$ is the phonon operator, $\bm{\hat{n}}_\lambda(\bm{q})$ the unit polarization vector of mode $\lambda$, and we defined the reduced mass $\mu_{S}^{-1}=M_\mathrm{TM}^{-1}+\left(3M_\mathrm{O}\right)^{-1}$ with ionic masses $M_\mathrm{O}$, $M_\mathrm{TM}$.
Following several experiments and \emph{ab initio} computations~\cite{Cowley1964,Harada1970,Vogt1988,Kozina2019}, we assumed that the soft FE mode can be approximated by the Slater mode where the TM atom moves against a rigid oxygen cage, $ M_\mathrm{TM}\bm{u}^{\mathrm{TM}}(\bm{q})=-3M_\mathrm{O}\bm{u}^{\mathrm{O}}(\bm{q})$ (see Appendix~\ref{app:el-ph}). Interband coupling terms have been neglected in Eq.~\eqref{eq:Helph}. 

Unlike the conventional gradient deformation potential, the Rashba coupling Eq.~\eqref{eq:glambda} stays finite at $\bm{q=0}$.
Furthermore, it selects the in-plane components of the phonon polarization vector $\bm{\hat{n}}_\lambda(\bm{q})$ ($E_u$ irreducible representation)
and thus it is only finite for any longitudinal or transverse mode with polarization in the $xy$ plane.
Additional terms to the Rashba coupling Eq.~\eqref{eq:Gamma1} allowed by symmetry~\cite{Sumita2020} are obtained for the electronic SOC states Eq.~\eqref{eq:soc} by considering spin-flip hopping processes $t_{xy}^{\uparrow \downarrow}$ in Eq.~\eqref{eq:Hu}: another $E_u$ term $k_z(\bm{\hat{z}}\times \bm{\sigma})$, and an $A_{2u}$ term $(\bm{k}\times \bm{\sigma})\cdot\bm{\hat{z}}$ that couples to the out-of-plane component of $\bm{\hat{n}}_\lambda(\bm{q})$. They each also carry, accordingly, a different induced hopping parameter $t_{xy}'$. For simplicity, we ignore these terms but note that they should be considered for a complete description of the Rashba coupling function of a SOC band.

Neglecting the crystal field on the phonon sector and assuming dispersion-less modes (i.e., staying in the $q\rightarrow 0$ limit) we obtain one of our main results which are simple expressions for the coupling Eq.~\eqref{eq:glambda} to the transverse optical (TO) modes (see Appendix~\ref{app:el-ph}):
\begin{eqnarray}
    && \mathfrak{g}_{TO1}(\bm{\hat k},\bm{\hat q})=g_{TO}\frac{i\left(\hat{k}_x\hat{q}_x+\hat{k}_y\hat{q}_y\right)}{\sqrt{1-\hat{q}_z^2}}
    \label{eq:gto1}
    \\
    && \mathfrak{g}_{TO2}(\bm{\hat k}, \bm{\hat q})=g_{TO}\frac{\hat{q}_z\left(\hat{k}_y\hat{q}_x-\hat{k}_x\hat{q}_y\right)}{\sqrt{1-\hat{q}_z^2}}
    \label{eq:gto2}
    \\
    &&g_{TO}=2 t'_{xy} \sqrt{\frac{\hbar}{2 \mu_S \omega_{TO}}} k_F a\equiv \alpha_{TO} k_F
    \label{eq:alphaTO}
\end{eqnarray}
We notice again that the Rashba parameter  $\alpha_{TO}$ and Rashba 
electron-phonon matrix element $g_{TO}$  are independent of the strength of $\lambda_{SO}$. On the other hand,  $g_{TO}$ is proportional to the Fermi wave vector $k_F$, and thus its magnitude will be accordingly reduced in very-low-density systems with a small Fermi energy.

\emph{Superconductivity.} We study now the superconducting solutions mediated by the Rashba coupling Eq.~\eqref{eq:glambda}.
We restrict the effective electron-electron interaction into the Cooper channel, $|\mathfrak{g}_{TO}(\bm{\hat{k}}+\bm{\hat{k}}',\bm{\hat{k}}-\bm{\hat{k}}')|^2 \mc{D}_{TO}$, with the matrix elements of the TO phonon sector given by Eqs.~\eqref{eq:gto1} and \eqref{eq:gto2} and the corresponding static phonon propagator $\mc{D}_{TO}=-\frac{2}{\omega_{TO}}$.

The resulting linearized gap equations at $T_c$ can be written in terms of a four-component vector $d_a$,  for the even-parity $\Delta_{\mathrm{even}}(\bm{\hat{k}})=d_0(\bm{\hat{k}}) i \sigma_y$ and odd-parity $\Delta_
 {\mathrm{odd}}(\bm{\hat{k}})=\bm{d}(\bm{\hat{k}})\cdot \bm{\sigma} i \sigma_y$  channels, 
 \begin{equation}
     d_a(\bm{\hat{k}})=N_F \frac{g_{TO}^2}{\omega_{TO}}\ln \left(\frac{1.13\omega_c}{k_B T_c}\right) \int \frac{d\bm{\hat{k}}'}{4\pi}L_a(\bm{\hat{k}},\bm{\hat{k}}')d_a(\bm{\hat{k}}')\\
   \label{eq:gapeq}
 \end{equation}
with $a=0$ for the singlet and $a=x,y,z$ for the triplet channel. 
 $N_F$ is the density of states (DOS) per spin at the Fermi energy, and $\omega_c$ is a characteristic cut-off frequency. 
 The interaction kernels are the following,
\begin{eqnarray}
    \label{eq:L0}
    &&L_0 (\bm{\hat{k}},\bm{\hat{k}}')=\frac{1}{2}\left[\ell (\bm{\hat{k}},\bm{\hat{k}}')+\ell (\bm{\hat{k}},-\bm{\hat{k}}')\right],
    \\
    \label{eq:Lz}
    &&L_z (\bm{\hat{k}},\bm{\hat{k}}')=\frac{1}{2}\left[\ell (\bm{\hat{k}},\bm{\hat{k}}')-\ell (\bm{\hat{k}},-\bm{\hat{k}}')\right],
    \\
    \label{eq:Lxy}
    &&L_{x,y} (\bm{\hat{k}},\bm{\hat{k}}')=-L_z (\bm{\hat{k}},\bm{\hat{k}}'),
\end{eqnarray}
with the effective interaction
\begin{equation}
    \ell (\bm{\hat{k}},\bm{\hat{k}}')=\frac{1+\bm{\hat{k}}\cdot\bm{\hat{k}}'}{1-\bm{\hat{k}}\cdot\bm{\hat{k}}'}\left(\hat{k}_z-\hat{k}_z'\right)^2.
    \label{eq:ell}
  \end{equation}
In the odd-parity channel, the sign change between the kernel for unequal pseudospin pairing  [Eq.~\eqref{eq:Lz}] and equal pseudospin pairing 
[Eq.~\eqref{eq:Lxy}] leads to  different critical temperatures $T_c$.
In other words, the present pairing interaction breaks the 
threefold degeneracy of the $T_{1u}$ $p$-wave channel that would be obtained in a cubic system~\cite{Sumita2020}.
Indeed, due to the $x,y$ orbital character of the electronic SOC band 
our interaction Eq.~\eqref{eq:ell} has cylindrical symmetry.

\begin{figure}
    \centering
    \includegraphics[width=\linewidth]{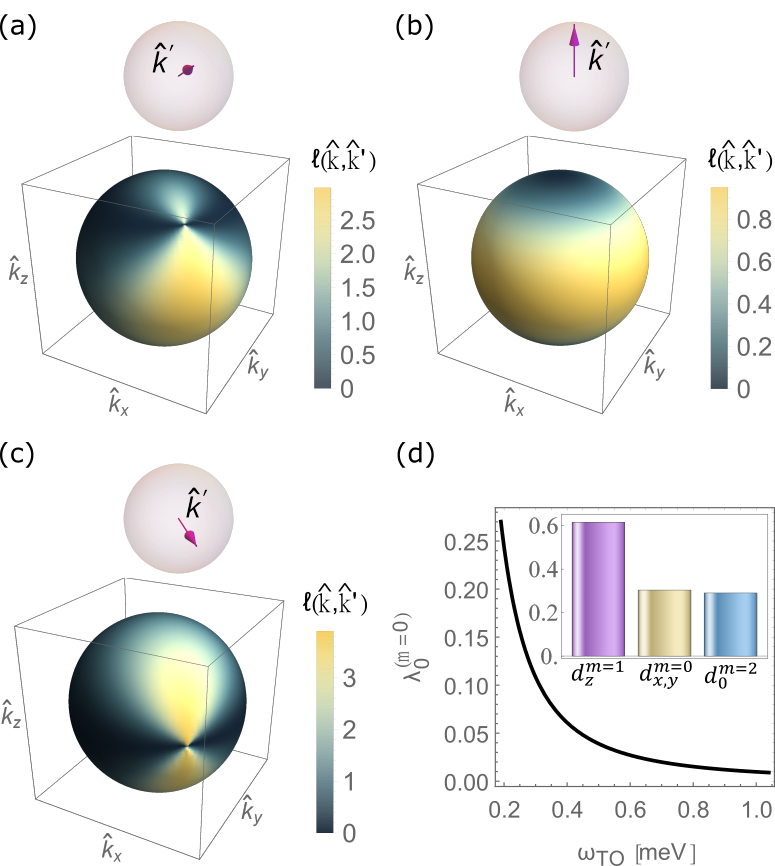}
    \caption{Angular dependence of the FE mode mediated interaction $\ell (\bm{\hat{k}},\bm{\hat{k}}')$ [Eq.~\eqref{eq:ell}] for $\bm{\hat{k}}'=(\theta',\varphi'=\pi/4)$ with (a) $\theta'=\pi/4$, (b)  $\theta'=0$, and (c) $\theta'=\pi/2$ , shown as magenta arrows in the inset. (d) BCS coupling constant [Eq.~\eqref{eq:lambda}] vs FE mode gap $\omega_{TO}$ for STO relevant parameters ($N_F=0.04$ eV$^{-1}$, $k_Fa=0.3$, and $t'_{xy}=108$ meV/{\AA}). Inset: $\lambda_a^{(m)}/\lambda_0^{(m=0)}$, eigenvalue ratio of sub-leading SC states 
    and $s$-wave channel.}
    \label{fig:sc}
\end{figure}

As can be seen from Figs.~\ref{fig:sc} (a)-(c), the momentum dependence on the Fermi surface is very strong.  It has large matrix elements when the momentum of the quasiparticles
$\bm{\hat{k}}$ and $\bm{\hat{k}}'$, is near the equator and the exchanged momentum is small [Fig.~\ref{fig:sc}(c)], while it is much weaker for quasiparticles moving along $z$ [Fig.~\ref{fig:sc}(b)]. 

Because of cylindrical symmetry, the effective interaction 
can be decomposed as an expansion of normalized real spherical harmonics $y_{lm}(\bm{\hat{k}})=\sqrt{4\pi}Y_{lm}(\bm{\hat{k}})$ of the same index $m$,  
  $ L_a (\bm{\hat{k}},\bm{\hat{k}}')=\sum_m \sum_{l,l'} v_{a,ll'}^{(m)} y_{lm}(\bm{\hat{k}})y_{l'm}(\bm{\hat{k}}')$. Thus the index $m$ corresponds to the irreducible representation of our cylindrical interaction, with even (odd) $l,l'$ for $a=0$ ($a=x,y,z$). 
The gap equation~\eqref{eq:gapeq} is then decoupled into orthogonal channels 
$d^{(m)}_a(\bm{\hat{k}})=\sum_l b^{(m)}_l y_{lm}(\bm{\hat{k}})$, each with an eigenvalue $\vartheta_a^{(m)}$ which enters the dimensionless BCS coupling constant
\begin{equation}
    \lambda_a^{(m)}= N_F \frac{g_{TO}^2}{\omega_{TO}}\vartheta_a^{(m)}= N_F \frac{\alpha^2_{TO} k_F^2}{\omega_{TO}}\vartheta_a^{(m)}
    \label{eq:lambda}
\end{equation}
with $k_B T_{c,a}^{(m)}=1.13\omega_c \exp \left[-1/\lambda_a^{(m)}\right]$.
Remarkably, despite the strong momentum dependence of the interaction, upon angular integration on $\bm{\hat{k}}'$ the result does not depend on $\bm{\hat{k}}$, i.e. $\int\frac{d\bm{\hat{k}}'}{4\pi} L_0(\bm{\hat{k}},\bm{\hat{k}}')=\frac{2}{3}$.
Consequently, the isotropic state is an “accidental” solution of the strongly momentum-dependent gap equation (\ref{eq:gapeq}), and it is indeed the leading pairing instability, $d_0^{(m=0)}(\bm{\hat{k}})=y_{00}$, with eigenvalue $\vartheta_0^{(m=0)}=\frac{2}{3}$.
The possibility of topological superconductivity of this state in interfaces with a magnetic field has been proposed~\cite{loder2015route}.

Notwithstanding the above trivial solution, the strong momentum dependence has important manifestations.
Traditionally, the only attractive channel of electron-phonon mediated SC is the even-parity $s$-wave.
Here, instead, the Rashba coupling yields attractive channels also in the odd-parity sector, in agreement with recent related studies~\cite{Kozii2015,Venderbos2016,Martin2017,Sumita2020}.
The inset of Fig.~\ref{fig:sc}(d) shows the eigenvalue ranking of the sub-leading solutions. 
Interestingly, the first two solutions belong to the odd-parity channel, $d_z^{(m=1)}$ and $d_{x,y}^{(m=0)}$, with a stronger pairing instability than the next attractive solution $d_0^{(m=2)}$, a $d$-wave singlet state. 
Thus, we obtain a $p$-wave solution as the first sub-leading state (see Appendix~\ref{app:channels}). We note that the role of the Coulomb interaction, neglected in this work, should also be considered when studying subleading SC channels. Because the strongest pair-breaking effect happens in the leading s-wave channel, the Coulomb interaction can further promote odd-parity solutions~\cite{Martin2017}.

\emph{ Numerical estimates for STO.}
We have performed  DFT frozen phonon computations and obtained an induced hopping $t'_{xy}=108$ meV/{\AA} for the lowest band of doped STO (see Appendix~\ref{app:dft} for details).
This value is three times larger than the rough estimate in Ref.~\cite{Ruhman2016}, believed to be unrealistically high at the time.
It is tempting to apply this theory to the very low doping region in which only one band of STO is filled (and multi-band effects can be safely neglected), with density $n\lesssim 1.05 \times 10^{18}$ cm$^{-3}$ and zero resistance below $T_c\approx 0.1$ K. Experimentally, the optical gap of the soft FE mode is around $1$ meV below 4 K~\cite{Shirane1969,Yamanaka2000} and remains soft in the doping range we consider~\cite{Bauerle1980}.
This results in a characteristic zero-point motion length of the Slater mode $l_S=\sqrt{\frac{\hbar}{2 \mu_S \omega_{TO}}}=0.56 a_0= 0.3 ${\AA}.
Thus, we get the following Rashba coupling [Eq.~\eqref{eq:alphaTO}]
\begin{equation}
    g_{TO}^{\mathrm{(STO)}}\approx 65 \mathrm{ meV} \cdot k_F a = \left( 254 \ \mathrm{ meV \ \AA} \right) \cdot k_F \equiv \alpha_{TO}^{\mathrm{(STO)}} k_F.
    \label{eq:alphaTOest}
\end{equation}
Remarkably, because here the orbital angular momentum is nearly unquenched, the estimated Rashba parameter per unit length $\alpha_{TO}^{\mathrm{(STO)}}/l_S$ is two orders of magnitude larger than previous estimates for finite displacements at LAO/STO interfaces~\cite{Khalsa2013,Zhong2013} (see Appendix~\ref{app:dft}).
For the quoted density  $k_Fa\approx 0.3$ [see Fig.~\ref{fig:Smodel}(a) in Appendix~\ref{app:dft}], and thus we obtain a moderately large Rashba coupling, $g_{TO}^{\mathrm{(STO)}} \approx 20$ meV.

However, one should be careful of attempting to explain superconductivity in this regime. In fact, contrary to other systems with similarly low density\cite{prakash2017evidence,bretz2019superconductivity}, the Meissner effect and bulk superconductivity have not been observed below densities of order $n\approx 10^{19}$ cm$^{-3}$~\cite{bretz2019superconductivity,Koonce1967,Collignon2017,collignon2019}.
Ignoring this fact leads to insurmountable problems for a conventional electron-phonon mechanism. Indeed, from specific heat measurements~\cite{McCalla2019}, the DOS per spin is $N_F\approx 0.04$ eV$^{-1}$ for  $n\approx  10^{18}$  cm$^{-3}$. In order to have a strong enough SC coupling $\lambda=N_F V_{\mathrm{eff}}\approx 0.2$ that sustains a $T_c\approx 0.1$ K, the pairing interaction should be $V_{\mathrm{eff}}\approx 5$ eV at low doping,
a value typical for a Coulomb repulsion and too large for a conventional electron-phonon mediated attraction.
Indeed, using the DFT estimated Rashba coupling Eq.~\eqref{eq:alphaTOest} the resulting pairing interaction is $V_{\mathrm{eff}}=\frac{2g_{TO}^2}{3\omega_{TO}}\approx 250$ meV in the s-wave channel, and together with the $N_F$ quoted above, the SC coupling we obtain is  $\lambda\approx  0.01$. Thus, not surprisingly, bulk superconductivity cannot be sustained in the low-density regime within the present mechanism. Increasing the density to the region where Meissner effect is observed ($n\approx 3 \times 10^{19}$ cm$^{-3}$,  $k_F a\approx 0.9$) and assuming SC is dominated by the lowest band 
[Eqs.~\eqref{eq:lambda} and \eqref{eq:alphaTOest}], we obtain a SC $s$-wave coupling $\lambda\approx 0.3$. Hence, in this regime, the Rashba-like mechanism can support a homogeneous SC state. 
The leading $s$-wave solution is consistent with the insensitivity of $T_c$ to non-magnetic defects~\cite{lin2015s}, assuming dominant intra-band scattering~\cite{yue2021}.

The zero resistance state observed in the low-doping one-band regime remains to be explained. While a detailed modeling is beyond our present scope,
we can offer the following scenario.  The stability of polar domains is very sensitive to local strains which may be influenced by 
the proliferation of tetragonal domains in the samples~\cite{Dec2004} and the type of doping~\cite{collignon2019,bretz2019superconductivity,yoon2021low}. This opens the possibility of having regions with a softer mode, 
which would locally increase $V_{\mathrm{eff}}$ and hence SC $\lambda$, and may lead to filamentary superconductivity~\cite{collignon2019,bretz2019superconductivity,Caprara2020,Leridon2020,szot2021filamentary}. Two very recent experiments have indeed connected an enhancement of $T_c$ to local correlated polar distortions~\cite{hameed2020ferroelectric,salmani2021}. 
Since we find $\lambda\propto \omega_{TO}^{-2}$ [see Fig.~\ref{fig:sc}(d)], a reduction by a factor of 3-5 of the local TO mode frequency leads to a strong enough local pairing interaction ($\lambda\approx 0.1 - 0.25$) to explain the observed SC in the one-band regime and in an inhomogeneous scenario (see Appendix~\ref{app:inhomogeneous} for a detailed discussion). 
Filamentary superconductivity has also been emphasized above bulk $T_c$ at high densities~\cite{collignon2019} and at interfaces\cite{Caprara2013} .

\emph{Conclusions.} 
We developed a microscopic model for a linear Rashba-like coupling between a soft polar mode and conduction electrons, relevant for systems close to a polar instability. The explicit form of the electron-phonon Hamiltonian was derived for a $|\frac{3}{2},\pm \frac{3}{2} \rangle$ SOC band with a Slater polar phonon, believed to be relevant in STO.
Indeed, the estimate of the bare linear Rashba-like coupling we obtained with the aid of \emph{ab initio} calculations for the lowest band of STO is found to be substantial, unlike previous estimates of the gradient-like coupling. 

Following a weak-coupling approach, we studied the even and odd-parity Cooper channels by considering the effective interaction arising from the Rashba-like deformation potential. 
We found that the leading s-wave channel is followed by an attractive $p$-wave state. 
Using the estimated bare Rashba coupling $g_{TO}$ for the lowest band of STO, 
we obtained a BCS coupling $\lambda$ for the $s$-wave solution  in the low- and intermediate-density regimes. 
While its value can only explain $T_c$ in a spatially uniform scenario at intermediate densities, for the dilute one-band regime it may explain it in the presence of inhomogeneities due to a local softening of the FE mode. Our approach can be extended to other incipient ferroelectrics and interfaces~\cite{reyren2007,han2014two,liu2021two}. 

\begin{acknowledgments}
We thank F. Mauri and C. Muzzi for useful discussions at the early and latest stages of this work.   
We acknowledge financial support from the Italian
MIUR through Projects No. PRIN 2017Z8TS5B, and No. 20207ZXT4Z and by
Regione Lazio (L. R. 13/08) under project SIMAP. 
M.N.G. is supported by the Marie Skłodowska-Curie
individual fellowship Grant Agreement SILVERPATH No. 893943. We acknowledge the CINECA award under the ISCRA initiative Grants No. HP10C72OM1 and No. HP10BV0TBS, for the availability of high-performance computing resources and support.
\end{acknowledgments}

\appendix

\section{Interorbital hopping}
\label{app:inter}

In this appendix we provide more details on the derivation of Eq.~\eqref{eq:tmunukq0} in the main text. In the presence of a finite polar displacement of the lattice new hopping channels are induced between different orbitals $\mu$ and $\nu$,  
\begin{equation}
    \mc{H}_u= \sum_{\bm{r} \bm{\delta} \mu \nu \sigma \eta=\pm}  t^{(\mu\nu)}_{\bm{r}\bm{\delta}} c^\dagger_{\bm{r}\mu\sigma} c_{\bm{r}+\eta\bm{\delta},\nu\sigma}+\mathrm{h.c.},
\end{equation}
where $\bm{\delta}=\bm{\hat{x}},\bm{\hat{y}},\bm{\hat{z}}$ is the vector parallel to the bond joining nearest neighbor TM atoms. Assuming that the hopping only depends on the displacement between the two TM atoms and the bridging O$_\delta$ atom involved in the process [see Fig.~\ref{fig:model}(b) in the main text], in the form $\bm{u}^{\mathrm{O}_\delta}_{\bm{r}+\frac{\bm{\delta}}{2}} -\bm{u}^{\mathrm{TM}}_{\bm{r}}-\bm{u}^{\mathrm{TM}}_{\bm{r}+\bm{\delta}}$, derivatives in real space are simply related,
\begin{equation}
\frac{\partial t^{(\mu\nu)}_{\bm{r}\bm{\delta}}}{\partial u^{\mathrm{O}_\mu}_{\nu,\bm{r}+\frac{\bm{\delta}}{2}}}=-\frac{\partial t^{(\mu\nu)}_{\bm{r}\bm{\delta}}}{\partial u^{\mathrm{TM}}_{\nu,\bm{r}}}=-\frac{\partial t^{(\mu\nu)}_{\bm{r}\bm{\delta}}}{\partial u^{\mathrm{TM}}_{\nu,\bm{r}+\bm{\delta}}}.
\label{eq:derivatives}
\end{equation}
The possible spin-conserving hopping processes between a pair of orbitals $\mu$ and $\nu$, allowed only for displacements perpendicular to the bond $\bm{\delta}$, are encoded in the following,
\begin{align}
    t^{(\mu\nu)}_{\bm{r}\bm{\delta}}=&t'_{\mu\nu}\left[\left(\bm{\hat\mu}\times\bm{\hat\nu}\right)\times\bm{\delta}\right]\nonumber\\
    &\cdot\left[\bm{u}^{\mathrm{O}_\delta}_{\bm{r}+\frac{\bm{\delta}}{2}} - (\bm{\hat\mu}\cdot\bm{\delta})\bm{u}^{\mathrm{TM}}_{\bm{r}}- (\bm{\hat\nu}\cdot\bm{\delta})\bm{u}^{\mathrm{TM}}_{\bm{r}+\bm{\delta}}\right]
\end{align}
where $\bm{\hat{\mu}},\bm{\hat{\nu}}=\bm{\hat{x}},\bm{\hat{y}},\bm{\hat{z}}$ is a unitary vector normal to the plane spanned by the $t_{2g}$ orbital $\mu, \nu$. In addition, we have neglected small deviations from cubic symmetry and taken the derivatives  Eq.~\eqref{eq:derivatives} to be independent of $\bm{\delta}$ for the pair of orbitals for simplicity, i.e. $t'_{\mu\nu}\equiv \frac{\partial t^{(\mu\nu)}_{\bm{r}\bm{\delta}}}{\partial u^{\mathrm{O}_\mu}_{\nu,\bm{r}+\frac{\bm{\delta}}{2}}}$. 
Finally, by Fourier expanding the ionic displacement
\begin{equation}
    \bm{u}^{\kappa}_{\bm{r}}=\sum_{\bm{q}}\bm{u}^{\kappa}(\bm{q}) e^{i\bm{q}\cdot\left(\bm{r}+\bm{\delta}_{\kappa}\right)}
\end{equation}
 of atom $\kappa$ at position $\bm{r}+\bm{\delta}_\kappa$, we arrive at the following expression for the induced hopping:
 \begin{widetext}
 \begin{equation}
    t_{\mu\nu}(\bm{k},\bm{q})=2it'_{\mu\nu}\sum_{\bm{\delta}}\left(\bm{\hat\mu}\times\bm{\hat\nu}\right)\times\bm{\delta}\cdot\left\{\bm{u}^{\mathrm{O}_\delta}(\bm{q})\sin k_\delta - \bm{u}^{\mathrm{TM}}(\bm{q})\left[ (\bm{\hat\mu}\cdot\bm{\delta})\sin\left(k_\delta-\frac{q_\delta}{2}\right)+(\bm{\hat\nu}\cdot\bm{\delta})\sin\left(k_\delta+\frac{q_\delta}{2}\right)\right]\right\}.
    \label{eq:tmunu}
\end{equation}
\end{widetext}
In the $\bm{k},\bm{q}\rightarrow 0$ limit, and under a rigid oxygen cage condition $\bm{u}^{\mathrm{O}_x}(\bm q)=\bm{u}^{\mathrm{O}_y}(\bm q)=\bm{u}^{\mathrm{O}_z}(\bm q)\equiv \bm{u}^{\mathrm{O}}(\bm q)$, Eq.~\eqref{eq:tmunu} becomes Eq. (2) in the main text.

\section{Electron-phonon Hamiltonian for a Slater mode}
\label{app:el-ph}

In order to get the electron-phonon Hamiltonian in Eq.~\eqref{eq:glambda}, we follow the usual procedure and quantize the displacements in Eq.~\eqref{eq:Gamma1} by introducing the polarization vectors $\bm{\hat{e}}^{\kappa}_{\lambda}$ of the normal mode $\lambda$ for each atom $\kappa$,
\begin{equation}
    \bm{u}^{\kappa}(\bm{q})=\sum_{\lambda}\sqrt{\frac{\hbar}{2 N M_\kappa \omega_{\bm{q}\lambda}}}\bm{\hat{e}}^{\kappa}_{\lambda}\left(a_{\bm{q}\lambda}+a^{\dagger}_{-\bm{q}\lambda}\right).
\end{equation}
Assuming that the polar mode is a Slater mode~\cite{1967Axe,Harada1970}, i.e. a mode which satisfies the rigid cage condition between the TM atom and the oxygen cage $\bm{u}^{\mathrm{TM}}(\bm{q})=-\frac{3M_\mathrm{O}}{M_\mathrm{TM}}\bm{u}^{\mathrm{O}}(\bm{q})$, we have the following relations for the polarization vectors, 
\begin{eqnarray}
     &&\frac{\bm{\hat{e}}^{\mathrm{O}}_{\lambda}}{\sqrt{M_{\mathrm{O}}}}=\frac{\sqrt{\mu_S}}{3M_{\mathrm{O}}}\bm{\hat{n}}_\lambda(\bm{q}),
     \\
     &&\frac{\bm{\hat{e}}^{\mathrm{TM}}_{\lambda}}{\sqrt{M_{\mathrm{TM}}}}=-\frac{\sqrt{\mu_S}}{M_{\mathrm{TM}}}\bm{\hat{n}}_\lambda(\bm{q}).
\end{eqnarray}
Here, $\bm{\hat{n}}_\lambda(\bm{q})$ is the unit polarization vector of the mode $\lambda$ and we have introduced the reduced mass of the Slater mode $\mu_{S}^{-1}=M_\mathrm{TM}^{-1}+\left(3M_\mathrm{O}\right)^{-1}$.
Substituting these two expressions one arrives at Eq.~\eqref{eq:glambda}. Following an equivalent procedure one can derive the Hamiltonian for a different polar mode, with a different set of relations between the atomic displacements $\bm{u}^{\kappa}(\bm{q})$. 

For the expressions in Eqs.~\eqref{eq:gto1} and \eqref{eq:gto2} we neglected any anisotropy in the phonons and took the polarization vectors in the direction of the unit vectors defining a spherical coordinate system. That is, we took the longitudinal optical (LO) mode along the radial direction in ${\bf q}$-space  ($\hat {\bf n}_{LO}\equiv \hat {\bf q}$)  and we took the two TO modes in the perpendicular direction.

\section{Subleading superconducting channels}
\label{app:channels}

In this appendix we give more details of the first three attractive subleading channels: $d_z^{(m=1)}$, $d_{x,y}^{(m=0)}$, and $d_0^{(m=2)}$, shown in the inset of Fig.~\ref{fig:sc}(d). The order parameters for these solutions take the following form, 
\begin{align}
    &d_z^{(m=1)} : &\bm{d}(\bm{\hat{k}})&=\upsilon_1 (0,0,\hat{k}_x) + \upsilon_2 (0,0,\hat{k}_y) 
   \label{eq:dzm1}    \\
    &d_{x,y}^{(m=0)} : &\bm{d}(\bm{\hat{k}})&=\upsilon_3 (f(\hat{k}_z),0,0)+\upsilon_4 (0,f(\hat{k}_z),0)    \\\nonumber
    & &f(\hat{k}_z)&= \hat{k}_z^3- 0.65\hat{k}_z -0.36\hat{k}_z^5    \\
    &d_{0}^{(m=2)} : &d_0(\bm{\hat{k}})&=\upsilon_5 (\hat{k}_x^2-\hat{k}_y^2)  +\upsilon_6 \hat{k}_x\hat{k}_y
     \label{eq:d0m2}
\end{align}
to leading order in the multipole expansion. Note that while the first subleading state $d_z^{(m=1)}$ is a conventional $p$-wave, in the case of $d_{x,y}^{(m=0)}$, the truncation had to be done at higher order. The coefficients $\upsilon_i$ multiply the degenerate gap functions at $T_c$ in our minimal model, and thus interestingly, nematic and chiral solutions are allowed. 
A consistent study for which a particular $\upsilon_i$ combination to expect below $T_c$ requires going beyond the linearized gap equation. These will generally depend on details of different compounds and it should be examined on a case-by-case basis.

\section{Estimate of the induced hopping parameter $t'_{xy}$}
\label{app:dft}

The estimation of the induced hopping parameter $t'_{xy}$ [Eq.~\eqref{eq:Gamma1}] for the lowest band of STO is done with the aid of first-principles computations. We performed calculations  within DFT using the projector augmented-wave (PAW) method as implemented in VASP~\cite{vasp1,vasp2} and the Perdew-Burke-Ernzerhof generalized gradient approximation revised for solids (PBEsol)~\cite{pbesol}. We considered the tetragonal phase of SrTiO$_3$ (space group $I4/mcm$), whose structure has been relaxed until forces were smaller than 1 meV/\AA, using a 6$\times$6$\times$6 mesh of $k$-points and a plane-wave cutoff of 520 eV. The optimized lattice parameters of the tetragonal cell (pseudocubic equivalents given in parentheses) are $a$=$b$=5.508 (3.895) \AA and $c$=7.845 (3.922) \AA, in excellent agreement with experimental data~\cite{xray}, with a deviation of at most 0.6$\%$. Band-structure calculations have been performed with the inclusion of spin-orbit coupling, both in the fully relaxed structure and for small atomic distortions corresponding to the Slater mode at fixed cell parameters.

\begin{figure}
    \centering
    \includegraphics[width=\linewidth]{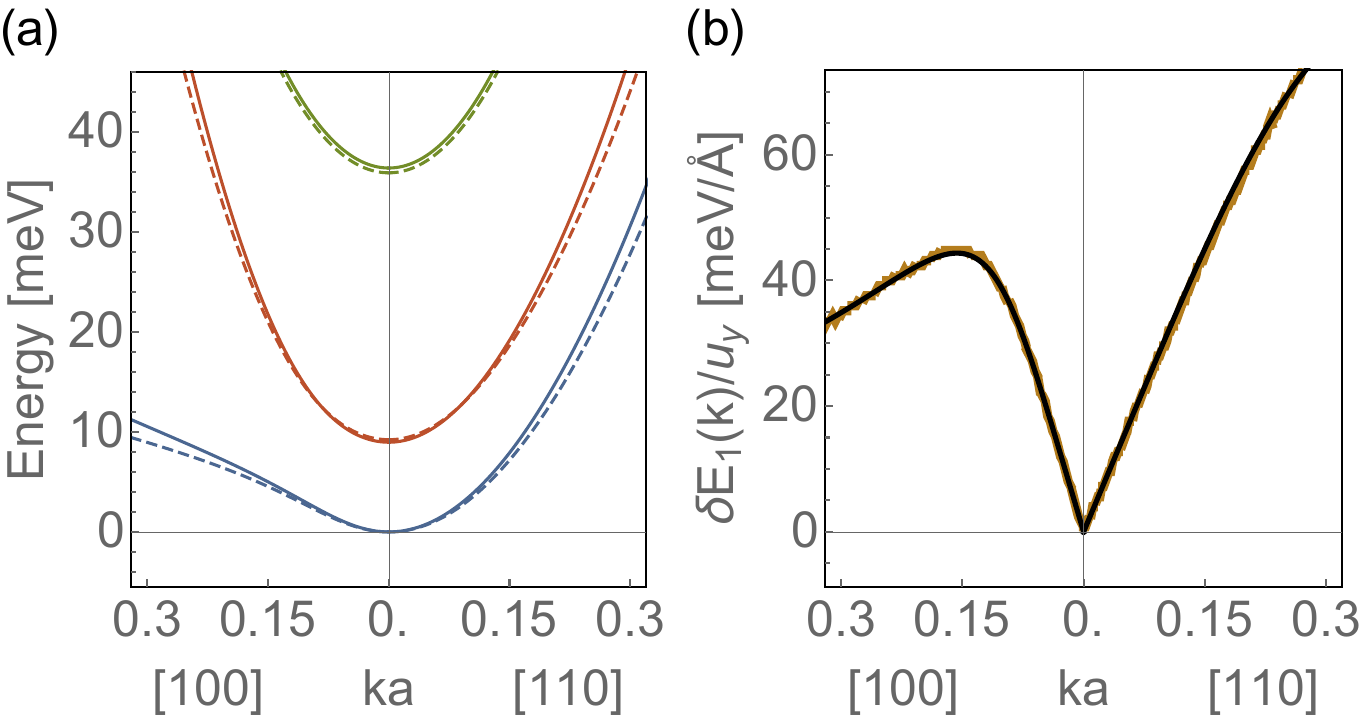}
    \caption{(a) Low energy band structure of STO as computed from DFT (dashed lines) and the tight-binding model (solid lines). (b) Rashba splitting of the lowest band $\delta E_1(\bm{k})$ in the presence finite polar displacement along $\bm{\hat{y}}$, $ u^{\mathrm{O}_x}_y-u^{\mathrm{Ti}}_y\equiv u_y=0.00125$ \AA. The thick solid orange line is the DFT frozen-phonon computation. The solid black line is obtained by diagonalizing the model Eq.~\eqref{eq:SH} with $t'_{xy}=108$ meV/\AA. }
    \label{fig:Smodel}
\end{figure}

The low energy electronic band structure of STO computed by DFT, shown in Fig.~\ref{fig:Smodel}(a) (dashed lines), consists of three doubly degenerate bands around the zone center.
A minimal tight-binding model with the $3d$ $t_{2g}$ orbitals of the Ti atom ($\mu=x,y,$ and $z$ as in the main text), extensively reported in earlier literature~\cite{Bistritzer2011,vanderMarel2011,Zhong2013,Gastiasoro2020review}, successfully describes the electronic band dispersion as shown by the solid lines in Fig.~\ref{fig:Smodel}(a). The model includes intraorbital and interorbital hopping terms $\mc{H}_0= \sum_{\bm{k}\mu \nu \sigma} \varepsilon_{\mu\nu}(\bm{k}) c^\dagger_{\mu\sigma}(\bm{k}) c_{\nu\sigma'}(\bm{k})$, with
\begin{align}
     \varepsilon_{\mu\mu}(\bm{k})=&-2t_1\left(\cos k_\alpha+ \cos k_\beta \right)-2 t_2 \cos k_\mu \nonumber\\
     &-4 t_3 \cos k_\alpha \cos k_\beta+(4t_1+2t_2+4t_3)\\
   \varepsilon_{xy}(\bm{k})=&-4 t_4 \sin k_x \sin k_y
 \end{align}
 Here, $\alpha,\beta \neq \mu$ and $\alpha\neq\beta$. The hopping parameters read $t_1=451$ meV, $t_2=40$ meV, $t_3=111$ meV and $t_4=27$ meV, similar to the ones found in Refs.~\cite{vanderMarel2011,Zhong2013}. The matrix defining the atomiclike spin-orbit coupling $\mc{H}_{SOC}$ and the tetragonal crystal field for the $t_{2g}$ orbitals $ \mc H_{AFD}$ in the $\left(x,y,z\right)\otimes\left(\uparrow,\downarrow\right)$ basis are
 \begin{equation}
 \xi\begin{pmatrix}
 0 & i & 0 & 0 & 0& -1 \\
 -i & 0 & 0 & 0 & 0& i\\
 0 & 0 & 0 & 1 & -i & 0\\
 0 & 0 & 1 & 0 & -i & 0\\
 0 & 0 & i & i & 0 & 0\\
 -1 & -i & 0 & 0 & 0 & 0 
 \end{pmatrix}
 +\Delta\begin{pmatrix}
 0 & 0 & 0 & 0 & 0 & 0 \\
 0 & 0 & 0 & 0 & 0 & 0\\
 0 & 0 & 1 & 0 & 0 & 0\\
 0 & 0 & 0 & 0 & 0 & 0\\
 0 & 0 & 0 & 0 & 0 & 0\\
 0 & 0 & 0 & 0 & 0 & 1 
 \end{pmatrix}
 \end{equation}
with $\xi=9.2$ meV describing the $3\xi=28$ meV gap opening at the zone center between the lower multiplet $j=3/2$ ($j_z=\pm 3/2,\pm 1/2$) and the higher doublet $j=1/2$  ($j_z=\pm 1/2$)  in the high-$T$ cubic symmetry. The antiferrodistortive (AFD) transition from cubic to tetragonal symmetry occurring at $105$ K is described by a simple splitting between the $x/y$ and the $z$ orbitals for $\Delta\neq0$.
This tetragonal crystal field term mixes states with $j_z=\pm 1/2$ (i.e. states with nonzero $\mu=z$ orbital character) and thus further splits the degeneracy of the lowest multiplet  $j=3/2$ at the zone center. 

A fit to the DFT band structure in the low-$T$ tetragonal phase [Fig.~\ref{fig:Smodel}(a)] gives $\Delta=17.7$ meV. The resulting lowest band at the zone center, which we focus on in this work, is the $|\frac{3}{2},\pm \frac{3}{2} \rangle$ doublet, involving only states with $x$ ($d_{yz}$) and $y$ ($d_{zx}$) orbital character. This may explain the difference with previous estimates for the Rashba coupling  in LAO/STO interfaces \cite{Zhong2013,Khalsa2013}, where the $2d$ confinement and the presence of the interface makes the $xy$ orbital the lowest-energy band. As argued in Ref. \cite{Djani2019}, the Rashba effect in perovskite oxides is strongly reduced in bands with a dominant $d_{xy}$ character, consistently with the results shown in Fig. 4 of Ref. \cite{Khalsa2013} where the Rashba spin splitting for higher-energy $yz$ and $zx$ bands is larger than the one found in the lowest-energy $xy$ band. 
 
In the presence of a polar displacement, an odd-parity interorbital hopping term $\mc{H}_u$ is included in the model [see Eqs.~\eqref{eq:Hu} and \eqref{eq:tmunu}], 
\begin{equation}
    \mc{H}=\mc{H}_0+\mc{H}_{SOC}+\mc H_{AFD}+\mc H_{u}
    \label{eq:SH}
\end{equation}
For the lowest band, formed by the $x$ and $y$ orbitals, only the term $t_{xy}(\bm{k},\bm{q})$ with $\mu=x$, $\nu=y$ in Eq.~\eqref{eq:tmunu} enters $\mc{H}_u$. 
In order to estimate the parameter $t'_{xy}$ in $\mc{H}_u$, we computed by first principles the reconstructed band structure of STO in the presence of a Slater frozen phonon, by displacing the Ti and O cage out of their equilibrium positions as explained in Appendix~\ref{app:el-ph}. The split of the lowest SOC band $\delta E_1(\bm{k})\equiv E_+(\bm{k})-E_-(\bm{k})$ normalized with the polar displacement amplitude $u_y=u^{\mathrm{O}_x}_y-u^{\mathrm{Ti}}_y$ is shown in Fig.~\ref{fig:Smodel} (b) (thick solid orange line). The same band split calculated from the band structure of our model Eq.~\eqref{eq:SH} with $t'_{xy}=108$ meV/\AA [black line in Fig.~\ref{fig:Smodel}(b)] gives an excellent fit to the DFT result. 
Indeed, as expected from the Rashba deformation potential expression [Eq. (4)], in the $\bm{k}\rightarrow 0$ limit, the lowest band is split in the conventional linear in $k$ Rashba form,
\begin{equation}
    \frac{\delta E_1(\bm{k})}{u_y}=4 t'_{xy} k_x a.
\end{equation}

\section{Inhomogeneous superconductivity in the low-doping regime}
\label{app:inhomogeneous}

The coupling estimation coming from our minimal model suggests that the mechanism involving the linear Rashba coupling to the FE fluctuations cannot support superconductivity on its own in the low-doping very-dilute regime of STO, assuming a spatial homogeneous scenario. On the other hand, accounting for a local softening of the FE mode due to inhomogeneities can explain several of the experimental observations in this very challenging regime. In this scenario, a proliferation of marginally stable modes may promote local superconductivity through a percolating path. The following facts point in this direction. 
First, the lowest doping superconducting dome is very sensitive to the kind of doping (it exists for oxygen doping and not for Nb doping) which suggests that it crucially depends on the nature of disorder~\cite{collignon2019}. 
Second, a careful experimental study in the single-band low-doping regime in oxygen reduced samples reported data consistent with an inhomogeneous superconducting state embedded within a homogeneous three-dimensional (3D) electron gas~\cite{bretz2019superconductivity}. In particular, the Meissner effect is absent, which indicates filamentary superconductivity~\cite{collignon2019}. This absence of a bulk manifestation of superconductivity contrasts with bismuth which shows bulk superconductivity at densities of the order of $10^{17}$ cm$^{–3}$ with a robust Meissner effect.\cite{prakash2017evidence}
Third, the influence of local correlated polar distortions enhancing superconductivity has been very recently highlighted in Ref.~\cite{salmani2021}. These works suggest the relevance of a scenario where strain fields around dislocations create regions characterized by a local enhanced softening of the TO mode. Previously, other works focusing on the insulating regime have shown that dislocations create strain gradients with polar regions stabilized close to the dislocation cores~\cite{gao2018atomic}. By continuity, polar critical regions should exist around the cores also in the weakly doped regime. Figure 6 of Ref.~\cite{hameed2020ferroelectric}  shows a semi-phenomenological computation of this effect.
Fourth, the relation between  dislocations, low-energy modes and superconductivity has been further emphasized very recently by Ref.~\cite{hameed2020ferroelectric} in which dislocations were intentionally created by plastic deformation. Plastically deformed samples show a proliferation of low-energy ferroelectric modes (as measured by polarized Raman) and enhanced $T_c$ of what the authors claim to be an inhomogeneous superconducting state. 
Interestingly, such marginally stable low temperature modes appear in a large class of disordered systems,  both in models and in experiments\cite{muller2015marginal,franz2015universal}. 

Therefore, while a spatially homogeneous superconducting scenario within our model seems unlikely in the low doping regime, we argue that an inhomogeneous superconducting state with the right order of magnitude of pairing temperature supported by this mechanism remains a strong candidate.

\bibliography{biblio}

\begin{thebibliography}{78}%
\makeatletter
\providecommand \@ifxundefined [1]{%
 \@ifx{#1\undefined}
}%
\providecommand \@ifnum [1]{%
 \ifnum #1\expandafter \@firstoftwo
 \else \expandafter \@secondoftwo
 \fi
}%
\providecommand \@ifx [1]{%
 \ifx #1\expandafter \@firstoftwo
 \else \expandafter \@secondoftwo
 \fi
}%
\providecommand \natexlab [1]{#1}%
\providecommand \enquote  [1]{``#1''}%
\providecommand \bibnamefont  [1]{#1}%
\providecommand \bibfnamefont [1]{#1}%
\providecommand \citenamefont [1]{#1}%
\providecommand \href@noop [0]{\@secondoftwo}%
\providecommand \href [0]{\begingroup \@sanitize@url \@href}%
\providecommand \@href[1]{\@@startlink{#1}\@@href}%
\providecommand \@@href[1]{\endgroup#1\@@endlink}%
\providecommand \@sanitize@url [0]{\catcode `\\12\catcode `\$12\catcode
  `\&12\catcode `\#12\catcode `\^12\catcode `\_12\catcode `\%12\relax}%
\providecommand \@@startlink[1]{}%
\providecommand \@@endlink[0]{}%
\providecommand \url  [0]{\begingroup\@sanitize@url \@url }%
\providecommand \@url [1]{\endgroup\@href {#1}{\urlprefix }}%
\providecommand \urlprefix  [0]{URL }%
\providecommand \Eprint [0]{\href }%
\providecommand \doibase [0]{https://doi.org/}%
\providecommand \selectlanguage [0]{\@gobble}%
\providecommand \bibinfo  [0]{\@secondoftwo}%
\providecommand \bibfield  [0]{\@secondoftwo}%
\providecommand \translation [1]{[#1]}%
\providecommand \BibitemOpen [0]{}%
\providecommand \bibitemStop [0]{}%
\providecommand \bibitemNoStop [0]{.\EOS\space}%
\providecommand \EOS [0]{\spacefactor3000\relax}%
\providecommand \BibitemShut  [1]{\csname bibitem#1\endcsname}%
\let\auto@bib@innerbib\@empty
\bibitem [{\citenamefont {Edge}\ \emph {et~al.}(2015)\citenamefont {Edge},
  \citenamefont {Kedem}, \citenamefont {Aschauer}, \citenamefont {Spaldin},\
  and\ \citenamefont {Balatsky}}]{Edge2015}%
  \BibitemOpen
  \bibfield  {author} {\bibinfo {author} {\bibfnamefont {J.~M.}\ \bibnamefont
  {Edge}}, \bibinfo {author} {\bibfnamefont {Y.}~\bibnamefont {Kedem}},
  \bibinfo {author} {\bibfnamefont {U.}~\bibnamefont {Aschauer}}, \bibinfo
  {author} {\bibfnamefont {N.~A.}\ \bibnamefont {Spaldin}},\ and\ \bibinfo
  {author} {\bibfnamefont {A.~V.}\ \bibnamefont {Balatsky}},\ }\bibfield
  {title} {\bibinfo {title} {Quantum critical origin of the superconducting
  dome in ${\mathrm{srtio}}_{3}$},\ }\href
  {https://doi.org/10.1103/PhysRevLett.115.247002} {\bibfield  {journal}
  {\bibinfo  {journal} {Phys. Rev. Lett.}\ }\textbf {\bibinfo {volume} {115}},\
  \bibinfo {pages} {247002} (\bibinfo {year} {2015})}\BibitemShut {NoStop}%
\bibitem [{\citenamefont {Kozii}\ \emph {et~al.}(2019)\citenamefont {Kozii},
  \citenamefont {Bi},\ and\ \citenamefont {Ruhman}}]{Kozii2019}%
  \BibitemOpen
  \bibfield  {author} {\bibinfo {author} {\bibfnamefont {V.}~\bibnamefont
  {Kozii}}, \bibinfo {author} {\bibfnamefont {Z.}~\bibnamefont {Bi}},\ and\
  \bibinfo {author} {\bibfnamefont {J.}~\bibnamefont {Ruhman}},\ }\bibfield
  {title} {\bibinfo {title} {Superconductivity near a ferroelectric quantum
  critical point in ultralow-density dirac materials},\ }\href
  {https://doi.org/10.1103/PhysRevX.9.031046} {\bibfield  {journal} {\bibinfo
  {journal} {Phys. Rev. X}\ }\textbf {\bibinfo {volume} {9}},\ \bibinfo {pages}
  {031046} (\bibinfo {year} {2019})}\BibitemShut {NoStop}%
\bibitem [{\citenamefont {Van Der~Marel}\ \emph {et~al.}(2019)\citenamefont
  {Van Der~Marel}, \citenamefont {Barantani},\ and\ \citenamefont
  {Rischau}}]{van2019possible}%
  \BibitemOpen
  \bibfield  {author} {\bibinfo {author} {\bibfnamefont {D.}~\bibnamefont {Van
  Der~Marel}}, \bibinfo {author} {\bibfnamefont {F.}~\bibnamefont
  {Barantani}},\ and\ \bibinfo {author} {\bibfnamefont {C.}~\bibnamefont
  {Rischau}},\ }\bibfield  {title} {\bibinfo {title} {Possible mechanism for
  superconductivity in doped srtio 3},\ }\href@noop {} {\bibfield  {journal}
  {\bibinfo  {journal} {Physical Review Research}\ }\textbf {\bibinfo {volume}
  {1}},\ \bibinfo {pages} {013003} (\bibinfo {year} {2019})}\BibitemShut
  {NoStop}%
\bibitem [{\citenamefont {Gastiasoro}\ \emph
  {et~al.}(2020{\natexlab{a}})\citenamefont {Gastiasoro}, \citenamefont
  {Trevisan},\ and\ \citenamefont {Fernandes}}]{Gastiasoro2020}%
  \BibitemOpen
  \bibfield  {author} {\bibinfo {author} {\bibfnamefont {M.~N.}\ \bibnamefont
  {Gastiasoro}}, \bibinfo {author} {\bibfnamefont {T.~V.}\ \bibnamefont
  {Trevisan}},\ and\ \bibinfo {author} {\bibfnamefont {R.~M.}\ \bibnamefont
  {Fernandes}},\ }\bibfield  {title} {\bibinfo {title} {Anisotropic
  superconductivity mediated by ferroelectric fluctuations in cubic systems
  with spin-orbit coupling},\ }\href
  {https://doi.org/10.1103/PhysRevB.101.174501} {\bibfield  {journal} {\bibinfo
   {journal} {Phys. Rev. B}\ }\textbf {\bibinfo {volume} {101}},\ \bibinfo
  {pages} {174501} (\bibinfo {year} {2020}{\natexlab{a}})}\BibitemShut
  {NoStop}%
\bibitem [{\citenamefont {Hameed}\ \emph {et~al.}(2022)\citenamefont {Hameed},
  \citenamefont {Pelc}, \citenamefont {Anderson}, \citenamefont {Klein},
  \citenamefont {Spieker}, \citenamefont {Yue}, \citenamefont {Das},
  \citenamefont {Ramberger}, \citenamefont {Lukas}, \citenamefont {Liu} \emph
  {et~al.}}]{hameed2020ferroelectric}%
  \BibitemOpen
  \bibfield  {author} {\bibinfo {author} {\bibfnamefont {S.}~\bibnamefont
  {Hameed}}, \bibinfo {author} {\bibfnamefont {D.}~\bibnamefont {Pelc}},
  \bibinfo {author} {\bibfnamefont {Z.~W.}\ \bibnamefont {Anderson}}, \bibinfo
  {author} {\bibfnamefont {A.}~\bibnamefont {Klein}}, \bibinfo {author}
  {\bibfnamefont {R.}~\bibnamefont {Spieker}}, \bibinfo {author} {\bibfnamefont
  {L.}~\bibnamefont {Yue}}, \bibinfo {author} {\bibfnamefont {B.}~\bibnamefont
  {Das}}, \bibinfo {author} {\bibfnamefont {J.}~\bibnamefont {Ramberger}},
  \bibinfo {author} {\bibfnamefont {M.}~\bibnamefont {Lukas}}, \bibinfo
  {author} {\bibfnamefont {Y.}~\bibnamefont {Liu}}, \emph {et~al.},\ }\bibfield
   {title} {\bibinfo {title} {Enhanced superconductivity and ferroelectric
  quantum criticality in plastically deformed strontium titanate},\ }\href@noop
  {} {\bibfield  {journal} {\bibinfo  {journal} {Nature Materials}\ }\textbf
  {\bibinfo {volume} {21}},\ \bibinfo {pages} {54} (\bibinfo {year}
  {2022})}\BibitemShut {NoStop}%
\bibitem [{\citenamefont {Yoon}\ \emph {et~al.}(2021)\citenamefont {Yoon},
  \citenamefont {Swartz}, \citenamefont {Harvey}, \citenamefont {Inoue},
  \citenamefont {Hikita}, \citenamefont {Yu}, \citenamefont {Chung},
  \citenamefont {Raghu},\ and\ \citenamefont {Hwang}}]{yoon2021low}%
  \BibitemOpen
  \bibfield  {author} {\bibinfo {author} {\bibfnamefont {H.}~\bibnamefont
  {Yoon}}, \bibinfo {author} {\bibfnamefont {A.~G.}\ \bibnamefont {Swartz}},
  \bibinfo {author} {\bibfnamefont {S.~P.}\ \bibnamefont {Harvey}}, \bibinfo
  {author} {\bibfnamefont {H.}~\bibnamefont {Inoue}}, \bibinfo {author}
  {\bibfnamefont {Y.}~\bibnamefont {Hikita}}, \bibinfo {author} {\bibfnamefont
  {Y.}~\bibnamefont {Yu}}, \bibinfo {author} {\bibfnamefont {S.~B.}\
  \bibnamefont {Chung}}, \bibinfo {author} {\bibfnamefont {S.}~\bibnamefont
  {Raghu}},\ and\ \bibinfo {author} {\bibfnamefont {H.~Y.}\ \bibnamefont
  {Hwang}},\ }\bibfield  {title} {\bibinfo {title} {Low-density
  superconductivity in srtio $ \_3 $ bounded by the adiabatic criterion},\
  }\href@noop {} {\bibfield  {journal} {\bibinfo  {journal} {arXiv preprint
  arXiv:2106.10802}\ } (\bibinfo {year} {2021})}\BibitemShut {NoStop}%
\bibitem [{\citenamefont {Fu}(2015)}]{Fu2015}%
  \BibitemOpen
  \bibfield  {author} {\bibinfo {author} {\bibfnamefont {L.}~\bibnamefont
  {Fu}},\ }\bibfield  {title} {\bibinfo {title} {Parity-breaking phases of
  spin-orbit-coupled metals with gyrotropic, ferroelectric, and multipolar
  orders},\ }\href {https://link.aps.org/doi/10.1103/PhysRevLett.115.026401}
  {\bibfield  {journal} {\bibinfo  {journal} {Phys. Rev. Lett.}\ }\textbf
  {\bibinfo {volume} {115}},\ \bibinfo {pages} {026401} (\bibinfo {year}
  {2015})}\BibitemShut {NoStop}%
\bibitem [{\citenamefont {Wu}\ and\ \citenamefont {Martin}(2017)}]{Martin2017}%
  \BibitemOpen
  \bibfield  {author} {\bibinfo {author} {\bibfnamefont {F.}~\bibnamefont
  {Wu}}\ and\ \bibinfo {author} {\bibfnamefont {I.}~\bibnamefont {Martin}},\
  }\bibfield  {title} {\bibinfo {title} {Nematic and chiral superconductivity
  induced by odd-parity fluctuations},\ }\href
  {https://doi.org/10.1103/PhysRevB.96.144504} {\bibfield  {journal} {\bibinfo
  {journal} {Phys. Rev. B}\ }\textbf {\bibinfo {volume} {96}},\ \bibinfo
  {pages} {144504} (\bibinfo {year} {2017})}\BibitemShut {NoStop}%
\bibitem [{\citenamefont {Schumann}\ \emph {et~al.}(2020)\citenamefont
  {Schumann}, \citenamefont {Galletti}, \citenamefont {Jeong}, \citenamefont
  {Ahadi}, \citenamefont {Strickland}, \citenamefont {Salmani-Rezaie},\ and\
  \citenamefont {Stemmer}}]{schumann2020possible}%
  \BibitemOpen
  \bibfield  {author} {\bibinfo {author} {\bibfnamefont {T.}~\bibnamefont
  {Schumann}}, \bibinfo {author} {\bibfnamefont {L.}~\bibnamefont {Galletti}},
  \bibinfo {author} {\bibfnamefont {H.}~\bibnamefont {Jeong}}, \bibinfo
  {author} {\bibfnamefont {K.}~\bibnamefont {Ahadi}}, \bibinfo {author}
  {\bibfnamefont {W.~M.}\ \bibnamefont {Strickland}}, \bibinfo {author}
  {\bibfnamefont {S.}~\bibnamefont {Salmani-Rezaie}},\ and\ \bibinfo {author}
  {\bibfnamefont {S.}~\bibnamefont {Stemmer}},\ }\bibfield  {title} {\bibinfo
  {title} {Possible signatures of mixed-parity superconductivity in doped polar
  srti o 3 films},\ }\href@noop {} {\bibfield  {journal} {\bibinfo  {journal}
  {Physical Review B}\ }\textbf {\bibinfo {volume} {101}},\ \bibinfo {pages}
  {100503} (\bibinfo {year} {2020})}\BibitemShut {NoStop}%
\bibitem [{\citenamefont {Sumita}\ and\ \citenamefont
  {Yanase}(2020)}]{Sumita2020}%
  \BibitemOpen
  \bibfield  {author} {\bibinfo {author} {\bibfnamefont {S.}~\bibnamefont
  {Sumita}}\ and\ \bibinfo {author} {\bibfnamefont {Y.}~\bibnamefont
  {Yanase}},\ }\bibfield  {title} {\bibinfo {title} {Superconductivity induced
  by fluctuations of momentum-based multipoles},\ }\href
  {https://doi.org/10.1103/PhysRevResearch.2.033225} {\bibfield  {journal}
  {\bibinfo  {journal} {Phys. Rev. Research}\ }\textbf {\bibinfo {volume}
  {2}},\ \bibinfo {pages} {033225} (\bibinfo {year} {2020})}\BibitemShut
  {NoStop}%
\bibitem [{\citenamefont {Zhou}\ \emph {et~al.}(2018)\citenamefont {Zhou},
  \citenamefont {Hellman},\ and\ \citenamefont {Bernardi}}]{zhou2018electron}%
  \BibitemOpen
  \bibfield  {author} {\bibinfo {author} {\bibfnamefont {J.-J.}\ \bibnamefont
  {Zhou}}, \bibinfo {author} {\bibfnamefont {O.}~\bibnamefont {Hellman}},\ and\
  \bibinfo {author} {\bibfnamefont {M.}~\bibnamefont {Bernardi}},\ }\bibfield
  {title} {\bibinfo {title} {Electron-phonon scattering in the presence of soft
  modes and electron mobility in srtio 3 perovskite from first principles},\
  }\href@noop {} {\bibfield  {journal} {\bibinfo  {journal} {Physical review
  letters}\ }\textbf {\bibinfo {volume} {121}},\ \bibinfo {pages} {226603}
  (\bibinfo {year} {2018})}\BibitemShut {NoStop}%
\bibitem [{\citenamefont {Wang}\ \emph {et~al.}(2019)\citenamefont {Wang},
  \citenamefont {Yang}, \citenamefont {Rischau}, \citenamefont {Xu},
  \citenamefont {Ren}, \citenamefont {Lorenz}, \citenamefont {Hemberger},
  \citenamefont {Lin},\ and\ \citenamefont {Behnia}}]{wang2019charge}%
  \BibitemOpen
  \bibfield  {author} {\bibinfo {author} {\bibfnamefont {J.}~\bibnamefont
  {Wang}}, \bibinfo {author} {\bibfnamefont {L.}~\bibnamefont {Yang}}, \bibinfo
  {author} {\bibfnamefont {C.~W.}\ \bibnamefont {Rischau}}, \bibinfo {author}
  {\bibfnamefont {Z.}~\bibnamefont {Xu}}, \bibinfo {author} {\bibfnamefont
  {Z.}~\bibnamefont {Ren}}, \bibinfo {author} {\bibfnamefont {T.}~\bibnamefont
  {Lorenz}}, \bibinfo {author} {\bibfnamefont {J.}~\bibnamefont {Hemberger}},
  \bibinfo {author} {\bibfnamefont {X.}~\bibnamefont {Lin}},\ and\ \bibinfo
  {author} {\bibfnamefont {K.}~\bibnamefont {Behnia}},\ }\bibfield  {title}
  {\bibinfo {title} {Charge transport in a polar metal},\ }\href@noop {}
  {\bibfield  {journal} {\bibinfo  {journal} {npj Quantum Materials}\ }\textbf
  {\bibinfo {volume} {4}},\ \bibinfo {pages} {1} (\bibinfo {year}
  {2019})}\BibitemShut {NoStop}%
\bibitem [{\citenamefont {Kumar}\ \emph {et~al.}(2021)\citenamefont {Kumar},
  \citenamefont {Yudson},\ and\ \citenamefont {Maslov}}]{kumar2021}%
  \BibitemOpen
  \bibfield  {author} {\bibinfo {author} {\bibfnamefont {A.}~\bibnamefont
  {Kumar}}, \bibinfo {author} {\bibfnamefont {V.~I.}\ \bibnamefont {Yudson}},\
  and\ \bibinfo {author} {\bibfnamefont {D.~L.}\ \bibnamefont {Maslov}},\
  }\bibfield  {title} {\bibinfo {title} {Quasiparticle and nonquasiparticle
  transport in doped quantum paraelectrics},\ }\href@noop {} {\bibfield
  {journal} {\bibinfo  {journal} {Physical Review Letters}\ }\textbf {\bibinfo
  {volume} {126}},\ \bibinfo {pages} {076601} (\bibinfo {year}
  {2021})}\BibitemShut {NoStop}%
\bibitem [{\citenamefont {Collignon}\ \emph {et~al.}(2021)\citenamefont
  {Collignon}, \citenamefont {Awashima}, \citenamefont {Lin}, \citenamefont
  {Rischau}, \citenamefont {Acheche}, \citenamefont {Vignolle}, \citenamefont
  {Proust}, \citenamefont {Fuseya}, \citenamefont {Behnia}, \citenamefont
  {Fauqu{\'e}} \emph {et~al.}}]{collignon2021quasi}%
  \BibitemOpen
  \bibfield  {author} {\bibinfo {author} {\bibfnamefont {C.}~\bibnamefont
  {Collignon}}, \bibinfo {author} {\bibfnamefont {Y.}~\bibnamefont {Awashima}},
  \bibinfo {author} {\bibfnamefont {X.}~\bibnamefont {Lin}}, \bibinfo {author}
  {\bibfnamefont {C.~W.}\ \bibnamefont {Rischau}}, \bibinfo {author}
  {\bibfnamefont {A.}~\bibnamefont {Acheche}}, \bibinfo {author} {\bibfnamefont
  {B.}~\bibnamefont {Vignolle}}, \bibinfo {author} {\bibfnamefont
  {C.}~\bibnamefont {Proust}}, \bibinfo {author} {\bibfnamefont
  {Y.}~\bibnamefont {Fuseya}}, \bibinfo {author} {\bibfnamefont
  {K.}~\bibnamefont {Behnia}}, \bibinfo {author} {\bibfnamefont
  {B.}~\bibnamefont {Fauqu{\'e}}}, \emph {et~al.},\ }\bibfield  {title}
  {\bibinfo {title} {Quasi-isotropic orbital magnetoresistance in lightly doped
  srtio 3},\ }\href@noop {} {\bibfield  {journal} {\bibinfo  {journal}
  {Physical Review Materials}\ }\textbf {\bibinfo {volume} {5}},\ \bibinfo
  {pages} {065002} (\bibinfo {year} {2021})}\BibitemShut {NoStop}%
\bibitem [{\citenamefont {Yue}\ \emph {et~al.}(2022)\citenamefont {Yue},
  \citenamefont {Ayino}, \citenamefont {Truttmann}, \citenamefont {Gastiasoro},
  \citenamefont {Persky}, \citenamefont {Khanukov}, \citenamefont {Lee},
  \citenamefont {Thoutam}, \citenamefont {Kalisky}, \citenamefont {Fernandes},
  \citenamefont {Pribiag},\ and\ \citenamefont {Jalan}}]{yue2021}%
  \BibitemOpen
  \bibfield  {author} {\bibinfo {author} {\bibfnamefont {J.}~\bibnamefont
  {Yue}}, \bibinfo {author} {\bibfnamefont {Y.}~\bibnamefont {Ayino}}, \bibinfo
  {author} {\bibfnamefont {T.~K.}\ \bibnamefont {Truttmann}}, \bibinfo {author}
  {\bibfnamefont {M.~N.}\ \bibnamefont {Gastiasoro}}, \bibinfo {author}
  {\bibfnamefont {E.}~\bibnamefont {Persky}}, \bibinfo {author} {\bibfnamefont
  {A.}~\bibnamefont {Khanukov}}, \bibinfo {author} {\bibfnamefont
  {D.}~\bibnamefont {Lee}}, \bibinfo {author} {\bibfnamefont {L.~R.}\
  \bibnamefont {Thoutam}}, \bibinfo {author} {\bibfnamefont {B.}~\bibnamefont
  {Kalisky}}, \bibinfo {author} {\bibfnamefont {R.~M.}\ \bibnamefont
  {Fernandes}}, \bibinfo {author} {\bibfnamefont {V.~S.}\ \bibnamefont
  {Pribiag}},\ and\ \bibinfo {author} {\bibfnamefont {B.}~\bibnamefont
  {Jalan}},\ }\bibfield  {title} {\bibinfo {title} {Anomalous transport in
  high-mobility superconducting srtio3 thin films},\ }\href
  {https://doi.org/10.1126/sciadv.abl5668} {\bibfield  {journal} {\bibinfo
  {journal} {Science Advances}\ }\textbf {\bibinfo {volume} {8}},\ \bibinfo
  {pages} {eabl5668} (\bibinfo {year} {2022})}\BibitemShut {NoStop}%
\bibitem [{\citenamefont {Volkov}\ and\ \citenamefont
  {Chandra}(2020)}]{Volkov2020}%
  \BibitemOpen
  \bibfield  {author} {\bibinfo {author} {\bibfnamefont {P.~A.}\ \bibnamefont
  {Volkov}}\ and\ \bibinfo {author} {\bibfnamefont {P.}~\bibnamefont
  {Chandra}},\ }\bibfield  {title} {\bibinfo {title} {Multiband quantum
  criticality of polar metals},\ }\href@noop {} {\bibfield  {journal} {\bibinfo
   {journal} {Physical review letters}\ }\textbf {\bibinfo {volume} {124}},\
  \bibinfo {pages} {237601} (\bibinfo {year} {2020})}\BibitemShut {NoStop}%
\bibitem [{\citenamefont {Stucky}\ \emph {et~al.}(2016)\citenamefont {Stucky},
  \citenamefont {Scheerer}, \citenamefont {Ren}, \citenamefont {Jaccard},
  \citenamefont {Poumirol}, \citenamefont {Barreteau}, \citenamefont
  {Giannini},\ and\ \citenamefont {van~der Marel}}]{Stucky2016}%
  \BibitemOpen
  \bibfield  {author} {\bibinfo {author} {\bibfnamefont {A.}~\bibnamefont
  {Stucky}}, \bibinfo {author} {\bibfnamefont {G.}~\bibnamefont {Scheerer}},
  \bibinfo {author} {\bibfnamefont {Z.}~\bibnamefont {Ren}}, \bibinfo {author}
  {\bibfnamefont {D.}~\bibnamefont {Jaccard}}, \bibinfo {author} {\bibfnamefont
  {J.-M.}\ \bibnamefont {Poumirol}}, \bibinfo {author} {\bibfnamefont
  {C.}~\bibnamefont {Barreteau}}, \bibinfo {author} {\bibfnamefont
  {E.}~\bibnamefont {Giannini}},\ and\ \bibinfo {author} {\bibfnamefont
  {D.}~\bibnamefont {van~der Marel}},\ }\bibfield  {title} {\bibinfo {title}
  {Isotope effect in superconducting n-doped srtio 3},\ }\href@noop {}
  {\bibfield  {journal} {\bibinfo  {journal} {Scientific reports}\ }\textbf
  {\bibinfo {volume} {6}},\ \bibinfo {pages} {37582} (\bibinfo {year}
  {2016})}\BibitemShut {NoStop}%
\bibitem [{\citenamefont {Rischau}\ \emph {et~al.}(2017)\citenamefont
  {Rischau}, \citenamefont {Lin}, \citenamefont {Grams}, \citenamefont {Finck},
  \citenamefont {Harms}, \citenamefont {Engelmayer}, \citenamefont {Lorenz},
  \citenamefont {Gallais}, \citenamefont {Fauque}, \citenamefont {Hemberger}
  \emph {et~al.}}]{Rischau2017}%
  \BibitemOpen
  \bibfield  {author} {\bibinfo {author} {\bibfnamefont {C.~W.}\ \bibnamefont
  {Rischau}}, \bibinfo {author} {\bibfnamefont {X.}~\bibnamefont {Lin}},
  \bibinfo {author} {\bibfnamefont {C.~P.}\ \bibnamefont {Grams}}, \bibinfo
  {author} {\bibfnamefont {D.}~\bibnamefont {Finck}}, \bibinfo {author}
  {\bibfnamefont {S.}~\bibnamefont {Harms}}, \bibinfo {author} {\bibfnamefont
  {J.}~\bibnamefont {Engelmayer}}, \bibinfo {author} {\bibfnamefont
  {T.}~\bibnamefont {Lorenz}}, \bibinfo {author} {\bibfnamefont
  {Y.}~\bibnamefont {Gallais}}, \bibinfo {author} {\bibfnamefont
  {B.}~\bibnamefont {Fauque}}, \bibinfo {author} {\bibfnamefont
  {J.}~\bibnamefont {Hemberger}}, \emph {et~al.},\ }\bibfield  {title}
  {\bibinfo {title} {A ferroelectric quantum phase transition inside the
  superconducting dome of sr 1- x ca x tio 3- $\delta$},\ }\href@noop {}
  {\bibfield  {journal} {\bibinfo  {journal} {Nature Physics}\ }\textbf
  {\bibinfo {volume} {13}},\ \bibinfo {pages} {643} (\bibinfo {year}
  {2017})}\BibitemShut {NoStop}%
\bibitem [{\citenamefont {Herrera}\ \emph {et~al.}(2019)\citenamefont
  {Herrera}, \citenamefont {Cerbin}, \citenamefont {Jayakody}, \citenamefont
  {Dunnett}, \citenamefont {Balatsky},\ and\ \citenamefont
  {Sochnikov}}]{Herrera2019}%
  \BibitemOpen
  \bibfield  {author} {\bibinfo {author} {\bibfnamefont {C.}~\bibnamefont
  {Herrera}}, \bibinfo {author} {\bibfnamefont {J.}~\bibnamefont {Cerbin}},
  \bibinfo {author} {\bibfnamefont {A.}~\bibnamefont {Jayakody}}, \bibinfo
  {author} {\bibfnamefont {K.}~\bibnamefont {Dunnett}}, \bibinfo {author}
  {\bibfnamefont {A.~V.}\ \bibnamefont {Balatsky}},\ and\ \bibinfo {author}
  {\bibfnamefont {I.}~\bibnamefont {Sochnikov}},\ }\bibfield  {title} {\bibinfo
  {title} {Strain-engineered interaction of quantum polar and superconducting
  phases},\ }\href@noop {} {\bibfield  {journal} {\bibinfo  {journal} {Physical
  Review Materials}\ }\textbf {\bibinfo {volume} {3}},\ \bibinfo {pages}
  {124801} (\bibinfo {year} {2019})}\BibitemShut {NoStop}%
\bibitem [{\citenamefont {Tomioka}\ \emph {et~al.}(2019)\citenamefont
  {Tomioka}, \citenamefont {Shirakawa}, \citenamefont {Shibuya},\ and\
  \citenamefont {Inoue}}]{Tomioka2019}%
  \BibitemOpen
  \bibfield  {author} {\bibinfo {author} {\bibfnamefont {Y.}~\bibnamefont
  {Tomioka}}, \bibinfo {author} {\bibfnamefont {N.}~\bibnamefont {Shirakawa}},
  \bibinfo {author} {\bibfnamefont {K.}~\bibnamefont {Shibuya}},\ and\ \bibinfo
  {author} {\bibfnamefont {I.~H.}\ \bibnamefont {Inoue}},\ }\bibfield  {title}
  {\bibinfo {title} {Enhanced superconductivity close to a non-magnetic quantum
  critical point in electron-doped strontium titanate},\ }\href@noop {}
  {\bibfield  {journal} {\bibinfo  {journal} {Nature Communications}\ }\textbf
  {\bibinfo {volume} {10}},\ \bibinfo {pages} {738} (\bibinfo {year}
  {2019})}\BibitemShut {NoStop}%
\bibitem [{\citenamefont {Ahadi}\ \emph {et~al.}(2019)\citenamefont {Ahadi},
  \citenamefont {Galletti}, \citenamefont {Li}, \citenamefont {Salmani-Rezaie},
  \citenamefont {Wu},\ and\ \citenamefont {Stemmer}}]{Ahadi2019}%
  \BibitemOpen
  \bibfield  {author} {\bibinfo {author} {\bibfnamefont {K.}~\bibnamefont
  {Ahadi}}, \bibinfo {author} {\bibfnamefont {L.}~\bibnamefont {Galletti}},
  \bibinfo {author} {\bibfnamefont {Y.}~\bibnamefont {Li}}, \bibinfo {author}
  {\bibfnamefont {S.}~\bibnamefont {Salmani-Rezaie}}, \bibinfo {author}
  {\bibfnamefont {W.}~\bibnamefont {Wu}},\ and\ \bibinfo {author}
  {\bibfnamefont {S.}~\bibnamefont {Stemmer}},\ }\bibfield  {title} {\bibinfo
  {title} {Enhancing superconductivity in srtio3 films with strain},\
  }\href@noop {} {\bibfield  {journal} {\bibinfo  {journal} {Science advances}\
  }\textbf {\bibinfo {volume} {5}},\ \bibinfo {pages} {eaaw0120} (\bibinfo
  {year} {2019})}\BibitemShut {NoStop}%
\bibitem [{\citenamefont {Russell}\ \emph {et~al.}(2019)\citenamefont
  {Russell}, \citenamefont {Ratcliff}, \citenamefont {Ahadi}, \citenamefont
  {Dong}, \citenamefont {Stemmer},\ and\ \citenamefont {Harter}}]{Harter2019}%
  \BibitemOpen
  \bibfield  {author} {\bibinfo {author} {\bibfnamefont {R.}~\bibnamefont
  {Russell}}, \bibinfo {author} {\bibfnamefont {N.}~\bibnamefont {Ratcliff}},
  \bibinfo {author} {\bibfnamefont {K.}~\bibnamefont {Ahadi}}, \bibinfo
  {author} {\bibfnamefont {L.}~\bibnamefont {Dong}}, \bibinfo {author}
  {\bibfnamefont {S.}~\bibnamefont {Stemmer}},\ and\ \bibinfo {author}
  {\bibfnamefont {J.~W.}\ \bibnamefont {Harter}},\ }\bibfield  {title}
  {\bibinfo {title} {Ferroelectric enhancement of superconductivity in
  compressively strained ${\mathrm{srtio}}_{3}$ films},\ }\href
  {https://doi.org/10.1103/PhysRevMaterials.3.091401} {\bibfield  {journal}
  {\bibinfo  {journal} {Phys. Rev. Materials}\ }\textbf {\bibinfo {volume}
  {3}},\ \bibinfo {pages} {091401} (\bibinfo {year} {2019})}\BibitemShut
  {NoStop}%
\bibitem [{\citenamefont {Enderlein}\ \emph {et~al.}(2020)\citenamefont
  {Enderlein}, \citenamefont {de~Oliveira}, \citenamefont {Tompsett},
  \citenamefont {Saitovitch}, \citenamefont {Saxena}, \citenamefont
  {Lonzarich},\ and\ \citenamefont {Rowley}}]{Enderlein2020}%
  \BibitemOpen
  \bibfield  {author} {\bibinfo {author} {\bibfnamefont {C.}~\bibnamefont
  {Enderlein}}, \bibinfo {author} {\bibfnamefont {J.~F.}\ \bibnamefont
  {de~Oliveira}}, \bibinfo {author} {\bibfnamefont {D.}~\bibnamefont
  {Tompsett}}, \bibinfo {author} {\bibfnamefont {E.~B.}\ \bibnamefont
  {Saitovitch}}, \bibinfo {author} {\bibfnamefont {S.}~\bibnamefont {Saxena}},
  \bibinfo {author} {\bibfnamefont {G.}~\bibnamefont {Lonzarich}},\ and\
  \bibinfo {author} {\bibfnamefont {S.}~\bibnamefont {Rowley}},\ }\bibfield
  {title} {\bibinfo {title} {Superconductivity mediated by polar modes in
  ferroelectric metals},\ }\href@noop {} {\bibfield  {journal} {\bibinfo
  {journal} {Nature communications}\ }\textbf {\bibinfo {volume} {11}},\
  \bibinfo {pages} {1} (\bibinfo {year} {2020})}\BibitemShut {NoStop}%
\bibitem [{\citenamefont {Franklin}\ \emph {et~al.}(2021)\citenamefont
  {Franklin}, \citenamefont {Xu}, \citenamefont {Davino}, \citenamefont
  {Mahabir}, \citenamefont {Balatsky}, \citenamefont {Aschauer},\ and\
  \citenamefont {Sochnikov}}]{franklin2021giant}%
  \BibitemOpen
  \bibfield  {author} {\bibinfo {author} {\bibfnamefont {J.}~\bibnamefont
  {Franklin}}, \bibinfo {author} {\bibfnamefont {B.}~\bibnamefont {Xu}},
  \bibinfo {author} {\bibfnamefont {D.}~\bibnamefont {Davino}}, \bibinfo
  {author} {\bibfnamefont {A.}~\bibnamefont {Mahabir}}, \bibinfo {author}
  {\bibfnamefont {A.~V.}\ \bibnamefont {Balatsky}}, \bibinfo {author}
  {\bibfnamefont {U.}~\bibnamefont {Aschauer}},\ and\ \bibinfo {author}
  {\bibfnamefont {I.}~\bibnamefont {Sochnikov}},\ }\bibfield  {title} {\bibinfo
  {title} {Giant gr{\"u}neisen parameter in a superconducting quantum
  paraelectric},\ }\href@noop {} {\bibfield  {journal} {\bibinfo  {journal}
  {Physical Review B}\ }\textbf {\bibinfo {volume} {103}},\ \bibinfo {pages}
  {214511} (\bibinfo {year} {2021})}\BibitemShut {NoStop}%
\bibitem [{\citenamefont {Salmani-Rezaie}\ \emph {et~al.}(2021)\citenamefont
  {Salmani-Rezaie}, \citenamefont {Jeong}, \citenamefont {Russell},
  \citenamefont {Harter},\ and\ \citenamefont {Stemmer}}]{salmani2021}%
  \BibitemOpen
  \bibfield  {author} {\bibinfo {author} {\bibfnamefont {S.}~\bibnamefont
  {Salmani-Rezaie}}, \bibinfo {author} {\bibfnamefont {H.}~\bibnamefont
  {Jeong}}, \bibinfo {author} {\bibfnamefont {R.}~\bibnamefont {Russell}},
  \bibinfo {author} {\bibfnamefont {J.~W.}\ \bibnamefont {Harter}},\ and\
  \bibinfo {author} {\bibfnamefont {S.}~\bibnamefont {Stemmer}},\ }\bibfield
  {title} {\bibinfo {title} {Role of locally polar regions in the
  superconductivity of srti o 3},\ }\href@noop {} {\bibfield  {journal}
  {\bibinfo  {journal} {Physical Review Materials}\ }\textbf {\bibinfo {volume}
  {5}},\ \bibinfo {pages} {104801} (\bibinfo {year} {2021})}\BibitemShut
  {NoStop}%
\bibitem [{\citenamefont {W\"olfle}\ and\ \citenamefont
  {Balatsky}(2018)}]{wolfle2018}%
  \BibitemOpen
  \bibfield  {author} {\bibinfo {author} {\bibfnamefont {P.}~\bibnamefont
  {W\"olfle}}\ and\ \bibinfo {author} {\bibfnamefont {A.~V.}\ \bibnamefont
  {Balatsky}},\ }\bibfield  {title} {\bibinfo {title} {Superconductivity at low
  density near a ferroelectric quantum critical point: Doped
  ${\mathrm{srtio}}_{3}$},\ }\href {https://doi.org/10.1103/PhysRevB.98.104505}
  {\bibfield  {journal} {\bibinfo  {journal} {Phys. Rev. B}\ }\textbf {\bibinfo
  {volume} {98}},\ \bibinfo {pages} {104505} (\bibinfo {year}
  {2018})}\BibitemShut {NoStop}%
\bibitem [{\citenamefont {W\"olfle}\ and\ \citenamefont
  {Balatsky}(2019)}]{wolfle2019reply}%
  \BibitemOpen
  \bibfield  {author} {\bibinfo {author} {\bibfnamefont {P.}~\bibnamefont
  {W\"olfle}}\ and\ \bibinfo {author} {\bibfnamefont {A.~V.}\ \bibnamefont
  {Balatsky}},\ }\bibfield  {title} {\bibinfo {title} {Reply to ``comment on
  `superconductivity at low density near a ferroelectric quantum critical
  point: Doped ${\mathrm{srtio}}_{3}$'''},\ }\href
  {https://doi.org/10.1103/PhysRevB.100.226502} {\bibfield  {journal} {\bibinfo
   {journal} {Phys. Rev. B}\ }\textbf {\bibinfo {volume} {100}},\ \bibinfo
  {pages} {226502} (\bibinfo {year} {2019})}\BibitemShut {NoStop}%
\bibitem [{\citenamefont {Arce-Gamboa}\ and\ \citenamefont
  {Guzm\'an-Verri}(2018)}]{Arce-Gamboa2018quantum}%
  \BibitemOpen
  \bibfield  {author} {\bibinfo {author} {\bibfnamefont {J.~R.}\ \bibnamefont
  {Arce-Gamboa}}\ and\ \bibinfo {author} {\bibfnamefont {G.~G.}\ \bibnamefont
  {Guzm\'an-Verri}},\ }\bibfield  {title} {\bibinfo {title} {Quantum
  ferroelectric instabilities in superconducting ${\mathrm{srtio}}_{3}$},\
  }\href {https://doi.org/10.1103/PhysRevMaterials.2.104804} {\bibfield
  {journal} {\bibinfo  {journal} {Phys. Rev. Materials}\ }\textbf {\bibinfo
  {volume} {2}},\ \bibinfo {pages} {104804} (\bibinfo {year}
  {2018})}\BibitemShut {NoStop}%
\bibitem [{\citenamefont {Kedem}(2018)}]{kedem2018novel}%
  \BibitemOpen
  \bibfield  {author} {\bibinfo {author} {\bibfnamefont {Y.}~\bibnamefont
  {Kedem}},\ }\bibfield  {title} {\bibinfo {title} {Novel pairing mechanism for
  superconductivity at a vanishing level of doping driven by critical
  ferroelectric modes},\ }\href {https://doi.org/10.1103/PhysRevB.98.220505}
  {\bibfield  {journal} {\bibinfo  {journal} {Phys. Rev. B}\ }\textbf {\bibinfo
  {volume} {98}},\ \bibinfo {pages} {220505} (\bibinfo {year}
  {2018})}\BibitemShut {NoStop}%
\bibitem [{\citenamefont {Gastiasoro}\ \emph
  {et~al.}(2020{\natexlab{b}})\citenamefont {Gastiasoro}, \citenamefont
  {Ruhman},\ and\ \citenamefont {Fernandes}}]{Gastiasoro2020review}%
  \BibitemOpen
  \bibfield  {author} {\bibinfo {author} {\bibfnamefont {M.~N.}\ \bibnamefont
  {Gastiasoro}}, \bibinfo {author} {\bibfnamefont {J.}~\bibnamefont {Ruhman}},\
  and\ \bibinfo {author} {\bibfnamefont {R.~M.}\ \bibnamefont {Fernandes}},\
  }\bibfield  {title} {\bibinfo {title} {Superconductivity in dilute srtio3: a
  review},\ }\href@noop {} {\bibfield  {journal} {\bibinfo  {journal} {Annals
  of Physics}\ }\textbf {\bibinfo {volume} {417}},\ \bibinfo {pages} {168107}
  (\bibinfo {year} {2020}{\natexlab{b}})}\BibitemShut {NoStop}%
\bibitem [{\citenamefont {Volkov}\ \emph {et~al.}(2021)\citenamefont {Volkov},
  \citenamefont {Chandra},\ and\ \citenamefont
  {Coleman}}]{volkov2021superconductivity}%
  \BibitemOpen
  \bibfield  {author} {\bibinfo {author} {\bibfnamefont {P.~A.}\ \bibnamefont
  {Volkov}}, \bibinfo {author} {\bibfnamefont {P.}~\bibnamefont {Chandra}},\
  and\ \bibinfo {author} {\bibfnamefont {P.}~\bibnamefont {Coleman}},\
  }\bibfield  {title} {\bibinfo {title} {Superconductivity from energy
  fluctuations in dilute quantum critical polar metals},\ }\href@noop {}
  {\bibfield  {journal} {\bibinfo  {journal} {arXiv preprint arXiv:2106.11295}\
  } (\bibinfo {year} {2021})}\BibitemShut {NoStop}%
\bibitem [{\citenamefont {Kiseliov}\ and\ \citenamefont
  {Feigel'man}(2021)}]{kiseliov2021theory}%
  \BibitemOpen
  \bibfield  {author} {\bibinfo {author} {\bibfnamefont {D.}~\bibnamefont
  {Kiseliov}}\ and\ \bibinfo {author} {\bibfnamefont {M.}~\bibnamefont
  {Feigel'man}},\ }\bibfield  {title} {\bibinfo {title} {Theory of
  superconductivity due to ngai's mechanism in lightly doped srtio3},\
  }\href@noop {} {\bibfield  {journal} {\bibinfo  {journal} {arXiv preprint
  arXiv:2106.09530}\ } (\bibinfo {year} {2021})}\BibitemShut {NoStop}%
\bibitem [{\citenamefont {Ruhman}\ and\ \citenamefont
  {Lee}(2019)}]{ruhman2019comment}%
  \BibitemOpen
  \bibfield  {author} {\bibinfo {author} {\bibfnamefont {J.}~\bibnamefont
  {Ruhman}}\ and\ \bibinfo {author} {\bibfnamefont {P.~A.}\ \bibnamefont
  {Lee}},\ }\bibfield  {title} {\bibinfo {title} {Comment on
  “superconductivity at low density near a ferroelectric quantum critical
  point: Doped srtio 3”},\ }\href@noop {} {\bibfield  {journal} {\bibinfo
  {journal} {Physical Review B}\ }\textbf {\bibinfo {volume} {100}},\ \bibinfo
  {pages} {226501} (\bibinfo {year} {2019})}\BibitemShut {NoStop}%
\bibitem [{\citenamefont {Ngai}(1974)}]{Ngai}%
  \BibitemOpen
  \bibfield  {author} {\bibinfo {author} {\bibfnamefont {K.~L.}\ \bibnamefont
  {Ngai}},\ }\bibfield  {title} {\bibinfo {title} {Two-phonon deformation
  potential and superconductivity in degenerate semiconductors},\ }\href
  {https://link.aps.org/doi/10.1103/PhysRevLett.32.215} {\bibfield  {journal}
  {\bibinfo  {journal} {Phys. Rev. Lett.}\ }\textbf {\bibinfo {volume} {32}},\
  \bibinfo {pages} {215} (\bibinfo {year} {1974})}\BibitemShut {NoStop}%
\bibitem [{\citenamefont {Takada}(1980)}]{Takada1980}%
  \BibitemOpen
  \bibfield  {author} {\bibinfo {author} {\bibfnamefont {Y.}~\bibnamefont
  {Takada}},\ }\bibfield  {title} {\bibinfo {title} {Theory of
  superconductivity in polar semiconductors and its application to n-type
  semiconducting srtio3},\ }\href@noop {} {\bibfield  {journal} {\bibinfo
  {journal} {Journal of the Physical Society of Japan}\ }\textbf {\bibinfo
  {volume} {49}},\ \bibinfo {pages} {1267} (\bibinfo {year}
  {1980})}\BibitemShut {NoStop}%
\bibitem [{\citenamefont {Ruhman}\ and\ \citenamefont
  {Lee}(2016)}]{Ruhman2016}%
  \BibitemOpen
  \bibfield  {author} {\bibinfo {author} {\bibfnamefont {J.}~\bibnamefont
  {Ruhman}}\ and\ \bibinfo {author} {\bibfnamefont {P.~A.}\ \bibnamefont
  {Lee}},\ }\bibfield  {title} {\bibinfo {title} {Superconductivity at very low
  density: The case of strontium titanate},\ }\href
  {https://doi.org/10.1103/PhysRevB.94.224515} {\bibfield  {journal} {\bibinfo
  {journal} {Phys. Rev. B}\ }\textbf {\bibinfo {volume} {94}},\ \bibinfo
  {pages} {224515} (\bibinfo {year} {2016})}\BibitemShut {NoStop}%
\bibitem [{\citenamefont {Kanasugi}\ and\ \citenamefont
  {Yanase}(2019)}]{Kanasugi2019}%
  \BibitemOpen
  \bibfield  {author} {\bibinfo {author} {\bibfnamefont {S.}~\bibnamefont
  {Kanasugi}}\ and\ \bibinfo {author} {\bibfnamefont {Y.}~\bibnamefont
  {Yanase}},\ }\bibfield  {title} {\bibinfo {title} {Multiorbital ferroelectric
  superconductivity in doped ${\mathrm{srtio}}_{3}$},\ }\href
  {https://doi.org/10.1103/PhysRevB.100.094504} {\bibfield  {journal} {\bibinfo
   {journal} {Phys. Rev. B}\ }\textbf {\bibinfo {volume} {100}},\ \bibinfo
  {pages} {094504} (\bibinfo {year} {2019})}\BibitemShut {NoStop}%
\bibitem [{\citenamefont {Petersen}\ and\ \citenamefont
  {Hedeg{\aa}rd}(2000)}]{Petersen2000}%
  \BibitemOpen
  \bibfield  {author} {\bibinfo {author} {\bibfnamefont {L.}~\bibnamefont
  {Petersen}}\ and\ \bibinfo {author} {\bibfnamefont {P.}~\bibnamefont
  {Hedeg{\aa}rd}},\ }\bibfield  {title} {\bibinfo {title} {A simple
  tight-binding model of spin--orbit splitting of sp-derived surface states},\
  }\href@noop {} {\bibfield  {journal} {\bibinfo  {journal} {Surface science}\
  }\textbf {\bibinfo {volume} {459}},\ \bibinfo {pages} {49} (\bibinfo {year}
  {2000})}\BibitemShut {NoStop}%
\bibitem [{\citenamefont {Khalsa}\ \emph {et~al.}(2013)\citenamefont {Khalsa},
  \citenamefont {Lee},\ and\ \citenamefont {MacDonald}}]{Khalsa2013}%
  \BibitemOpen
  \bibfield  {author} {\bibinfo {author} {\bibfnamefont {G.}~\bibnamefont
  {Khalsa}}, \bibinfo {author} {\bibfnamefont {B.}~\bibnamefont {Lee}},\ and\
  \bibinfo {author} {\bibfnamefont {A.~H.}\ \bibnamefont {MacDonald}},\
  }\bibfield  {title} {\bibinfo {title} {Theory of t 2 g electron-gas rashba
  interactions},\ }\href@noop {} {\bibfield  {journal} {\bibinfo  {journal}
  {Physical Review B}\ }\textbf {\bibinfo {volume} {88}},\ \bibinfo {pages}
  {041302} (\bibinfo {year} {2013})}\BibitemShut {NoStop}%
\bibitem [{\citenamefont {Zhong}\ \emph {et~al.}(2013)\citenamefont {Zhong},
  \citenamefont {T\'oth},\ and\ \citenamefont {Held}}]{Zhong2013}%
  \BibitemOpen
  \bibfield  {author} {\bibinfo {author} {\bibfnamefont {Z.}~\bibnamefont
  {Zhong}}, \bibinfo {author} {\bibfnamefont {A.}~\bibnamefont {T\'oth}},\ and\
  \bibinfo {author} {\bibfnamefont {K.}~\bibnamefont {Held}},\ }\bibfield
  {title} {\bibinfo {title} {Theory of spin-orbit coupling at
  laalo${}_{3}$/srtio${}_{3}$ interfaces and srtio${}_{3}$ surfaces},\ }\href
  {https://doi.org/10.1103/PhysRevB.87.161102} {\bibfield  {journal} {\bibinfo
  {journal} {Phys. Rev. B}\ }\textbf {\bibinfo {volume} {87}},\ \bibinfo
  {pages} {161102} (\bibinfo {year} {2013})}\BibitemShut {NoStop}%
\bibitem [{\citenamefont {Kozii}\ and\ \citenamefont {Fu}(2015)}]{Kozii2015}%
  \BibitemOpen
  \bibfield  {author} {\bibinfo {author} {\bibfnamefont {V.}~\bibnamefont
  {Kozii}}\ and\ \bibinfo {author} {\bibfnamefont {L.}~\bibnamefont {Fu}},\
  }\bibfield  {title} {\bibinfo {title} {Odd-parity superconductivity in the
  vicinity of inversion symmetry breaking in spin-orbit-coupled systems},\
  }\href {https://doi.org/10.1103/PhysRevLett.115.207002} {\bibfield  {journal}
  {\bibinfo  {journal} {Phys. Rev. Lett.}\ }\textbf {\bibinfo {volume} {115}},\
  \bibinfo {pages} {207002} (\bibinfo {year} {2015})}\BibitemShut {NoStop}%
\bibitem [{\citenamefont {Kanasugi}\ and\ \citenamefont
  {Yanase}(2018)}]{Kanasugi2018}%
  \BibitemOpen
  \bibfield  {author} {\bibinfo {author} {\bibfnamefont {S.}~\bibnamefont
  {Kanasugi}}\ and\ \bibinfo {author} {\bibfnamefont {Y.}~\bibnamefont
  {Yanase}},\ }\bibfield  {title} {\bibinfo {title} {Spin-orbit-coupled
  ferroelectric superconductivity},\ }\href
  {https://doi.org/10.1103/PhysRevB.98.024521} {\bibfield  {journal} {\bibinfo
  {journal} {Phys. Rev. B}\ }\textbf {\bibinfo {volume} {98}},\ \bibinfo
  {pages} {024521} (\bibinfo {year} {2018})}\BibitemShut {NoStop}%
\bibitem [{\citenamefont {Bistritzer}\ \emph {et~al.}(2011)\citenamefont
  {Bistritzer}, \citenamefont {Khalsa},\ and\ \citenamefont
  {MacDonald}}]{Bistritzer2011}%
  \BibitemOpen
  \bibfield  {author} {\bibinfo {author} {\bibfnamefont {R.}~\bibnamefont
  {Bistritzer}}, \bibinfo {author} {\bibfnamefont {G.}~\bibnamefont {Khalsa}},\
  and\ \bibinfo {author} {\bibfnamefont {A.~H.}\ \bibnamefont {MacDonald}},\
  }\bibfield  {title} {\bibinfo {title} {Electronic structure of doped
  ${d}^{0}$ perovskite semiconductors},\ }\href
  {https://doi.org/10.1103/PhysRevB.83.115114} {\bibfield  {journal} {\bibinfo
  {journal} {Phys. Rev. B}\ }\textbf {\bibinfo {volume} {83}},\ \bibinfo
  {pages} {115114} (\bibinfo {year} {2011})}\BibitemShut {NoStop}%
\bibitem [{\citenamefont {Djani}\ \emph {et~al.}(2019)\citenamefont {Djani},
  \citenamefont {Garcia-Castro}, \citenamefont {Tong}, \citenamefont {Barone},
  \citenamefont {Bousquet}, \citenamefont {Picozzi},\ and\ \citenamefont
  {Ghosez}}]{Djani2019}%
  \BibitemOpen
  \bibfield  {author} {\bibinfo {author} {\bibfnamefont {H.}~\bibnamefont
  {Djani}}, \bibinfo {author} {\bibfnamefont {A.~C.}\ \bibnamefont
  {Garcia-Castro}}, \bibinfo {author} {\bibfnamefont {W.-Y.}\ \bibnamefont
  {Tong}}, \bibinfo {author} {\bibfnamefont {P.}~\bibnamefont {Barone}},
  \bibinfo {author} {\bibfnamefont {E.}~\bibnamefont {Bousquet}}, \bibinfo
  {author} {\bibfnamefont {S.}~\bibnamefont {Picozzi}},\ and\ \bibinfo {author}
  {\bibfnamefont {P.}~\bibnamefont {Ghosez}},\ }\bibfield  {title} {\bibinfo
  {title} {Rationalizing and engineering rashba spin-splitting in ferroelectric
  oxides},\ }\href@noop {} {\bibfield  {journal} {\bibinfo  {journal} {npj
  Quantum Materials}\ }\textbf {\bibinfo {volume} {4}},\ \bibinfo {pages} {1}
  (\bibinfo {year} {2019})}\BibitemShut {NoStop}%
\bibitem [{\citenamefont {Slater}\ and\ \citenamefont {Koster}(1954)}]{slater}%
  \BibitemOpen
  \bibfield  {author} {\bibinfo {author} {\bibfnamefont {J.~C.}\ \bibnamefont
  {Slater}}\ and\ \bibinfo {author} {\bibfnamefont {G.~F.}\ \bibnamefont
  {Koster}},\ }\bibfield  {title} {\bibinfo {title} {Simplified lcao method for
  the periodic potential problem},\ }\href@noop {} {\bibfield  {journal}
  {\bibinfo  {journal} {Physical Review}\ }\textbf {\bibinfo {volume} {94}},\
  \bibinfo {pages} {1498} (\bibinfo {year} {1954})}\BibitemShut {NoStop}%
\bibitem [{\citenamefont {Cowley}(1964)}]{Cowley1964}%
  \BibitemOpen
  \bibfield  {author} {\bibinfo {author} {\bibfnamefont {R.~A.}\ \bibnamefont
  {Cowley}},\ }\bibfield  {title} {\bibinfo {title} {Lattice dynamics and phase
  transitions of strontium titanate},\ }\href
  {https://doi.org/10.1103/PhysRev.134.A981} {\bibfield  {journal} {\bibinfo
  {journal} {Phys. Rev.}\ }\textbf {\bibinfo {volume} {134}},\ \bibinfo {pages}
  {A981} (\bibinfo {year} {1964})}\BibitemShut {NoStop}%
\bibitem [{\citenamefont {Harada}\ \emph {et~al.}(1970)\citenamefont {Harada},
  \citenamefont {Axe},\ and\ \citenamefont {Shirane}}]{Harada1970}%
  \BibitemOpen
  \bibfield  {author} {\bibinfo {author} {\bibfnamefont {J.}~\bibnamefont
  {Harada}}, \bibinfo {author} {\bibfnamefont {J.}~\bibnamefont {Axe}},\ and\
  \bibinfo {author} {\bibfnamefont {G.}~\bibnamefont {Shirane}},\ }\bibfield
  {title} {\bibinfo {title} {Determination of the normal vibrational
  displacements in several perovskites by inelastic neutron scattering},\
  }\href@noop {} {\bibfield  {journal} {\bibinfo  {journal} {Acta
  Crystallographica Section A: Crystal Physics, Diffraction, Theoretical and
  General Crystallography}\ }\textbf {\bibinfo {volume} {26}},\ \bibinfo
  {pages} {608} (\bibinfo {year} {1970})}\BibitemShut {NoStop}%
\bibitem [{\citenamefont {Vogt}(1988)}]{Vogt1988}%
  \BibitemOpen
  \bibfield  {author} {\bibinfo {author} {\bibfnamefont {H.}~\bibnamefont
  {Vogt}},\ }\bibfield  {title} {\bibinfo {title} {Hyper-raman tensors of the
  zone-center optical phonons in srtio 3 and ktao 3},\ }\href@noop {}
  {\bibfield  {journal} {\bibinfo  {journal} {Physical Review B}\ }\textbf
  {\bibinfo {volume} {38}},\ \bibinfo {pages} {5699} (\bibinfo {year}
  {1988})}\BibitemShut {NoStop}%
\bibitem [{\citenamefont {Kozina}\ \emph {et~al.}(2019)\citenamefont {Kozina},
  \citenamefont {Fechner}, \citenamefont {Marsik}, \citenamefont {van Driel},
  \citenamefont {Glownia}, \citenamefont {Bernhard}, \citenamefont {Radovic},
  \citenamefont {Zhu}, \citenamefont {Bonetti}, \citenamefont {Staub} \emph
  {et~al.}}]{Kozina2019}%
  \BibitemOpen
  \bibfield  {author} {\bibinfo {author} {\bibfnamefont {M.}~\bibnamefont
  {Kozina}}, \bibinfo {author} {\bibfnamefont {M.}~\bibnamefont {Fechner}},
  \bibinfo {author} {\bibfnamefont {P.}~\bibnamefont {Marsik}}, \bibinfo
  {author} {\bibfnamefont {T.}~\bibnamefont {van Driel}}, \bibinfo {author}
  {\bibfnamefont {J.~M.}\ \bibnamefont {Glownia}}, \bibinfo {author}
  {\bibfnamefont {C.}~\bibnamefont {Bernhard}}, \bibinfo {author}
  {\bibfnamefont {M.}~\bibnamefont {Radovic}}, \bibinfo {author} {\bibfnamefont
  {D.}~\bibnamefont {Zhu}}, \bibinfo {author} {\bibfnamefont {S.}~\bibnamefont
  {Bonetti}}, \bibinfo {author} {\bibfnamefont {U.}~\bibnamefont {Staub}},
  \emph {et~al.},\ }\bibfield  {title} {\bibinfo {title} {Terahertz-driven
  phonon upconversion in srtio 3},\ }\href@noop {} {\bibfield  {journal}
  {\bibinfo  {journal} {Nature Physics}\ }\textbf {\bibinfo {volume} {15}},\
  \bibinfo {pages} {387} (\bibinfo {year} {2019})}\BibitemShut {NoStop}%
\bibitem [{\citenamefont {Loder}\ \emph {et~al.}(2015)\citenamefont {Loder},
  \citenamefont {Kampf},\ and\ \citenamefont {Kopp}}]{loder2015route}%
  \BibitemOpen
  \bibfield  {author} {\bibinfo {author} {\bibfnamefont {F.}~\bibnamefont
  {Loder}}, \bibinfo {author} {\bibfnamefont {A.~P.}\ \bibnamefont {Kampf}},\
  and\ \bibinfo {author} {\bibfnamefont {T.}~\bibnamefont {Kopp}},\ }\bibfield
  {title} {\bibinfo {title} {Route to topological superconductivity via
  magnetic field rotation},\ }\href@noop {} {\bibfield  {journal} {\bibinfo
  {journal} {Scientific reports}\ }\textbf {\bibinfo {volume} {5}},\ \bibinfo
  {pages} {1} (\bibinfo {year} {2015})}\BibitemShut {NoStop}%
\bibitem [{\citenamefont {Venderbos}\ \emph {et~al.}(2016)\citenamefont
  {Venderbos}, \citenamefont {Kozii},\ and\ \citenamefont
  {Fu}}]{Venderbos2016}%
  \BibitemOpen
  \bibfield  {author} {\bibinfo {author} {\bibfnamefont {J.~W.~F.}\
  \bibnamefont {Venderbos}}, \bibinfo {author} {\bibfnamefont {V.}~\bibnamefont
  {Kozii}},\ and\ \bibinfo {author} {\bibfnamefont {L.}~\bibnamefont {Fu}},\
  }\bibfield  {title} {\bibinfo {title} {Odd-parity superconductors with
  two-component order parameters: Nematic and chiral, full gap, and majorana
  node},\ }\href {https://doi.org/10.1103/PhysRevB.94.180504} {\bibfield
  {journal} {\bibinfo  {journal} {Phys. Rev. B}\ }\textbf {\bibinfo {volume}
  {94}},\ \bibinfo {pages} {180504} (\bibinfo {year} {2016})}\BibitemShut
  {NoStop}%
\bibitem [{\citenamefont {Shirane}\ and\ \citenamefont
  {Yamada}(1969)}]{Shirane1969}%
  \BibitemOpen
  \bibfield  {author} {\bibinfo {author} {\bibfnamefont {G.}~\bibnamefont
  {Shirane}}\ and\ \bibinfo {author} {\bibfnamefont {Y.}~\bibnamefont
  {Yamada}},\ }\bibfield  {title} {\bibinfo {title} {Lattice-dynamical study of
  the 110\ifmmode^\circ\else\textdegree\fi{}k phase transition in
  srti${\mathrm{o}}_{3}$},\ }\href
  {https://link.aps.org/doi/10.1103/PhysRev.177.858} {\bibfield  {journal}
  {\bibinfo  {journal} {Phys. Rev.}\ }\textbf {\bibinfo {volume} {177}},\
  \bibinfo {pages} {858} (\bibinfo {year} {1969})}\BibitemShut {NoStop}%
\bibitem [{\citenamefont {Yamanaka}\ \emph {et~al.}(2000)\citenamefont
  {Yamanaka}, \citenamefont {Kataoka}, \citenamefont {Inaba}, \citenamefont
  {Inoue}, \citenamefont {Hehlen},\ and\ \citenamefont
  {Courtens}}]{Yamanaka2000}%
  \BibitemOpen
  \bibfield  {author} {\bibinfo {author} {\bibfnamefont {A.}~\bibnamefont
  {Yamanaka}}, \bibinfo {author} {\bibfnamefont {M.}~\bibnamefont {Kataoka}},
  \bibinfo {author} {\bibfnamefont {Y.}~\bibnamefont {Inaba}}, \bibinfo
  {author} {\bibfnamefont {K.}~\bibnamefont {Inoue}}, \bibinfo {author}
  {\bibfnamefont {B.}~\bibnamefont {Hehlen}},\ and\ \bibinfo {author}
  {\bibfnamefont {E.}~\bibnamefont {Courtens}},\ }\bibfield  {title} {\bibinfo
  {title} {Evidence for competing orderings in strontium titanate from
  hyper-raman scattering spectroscopy},\ }\href@noop {} {\bibfield  {journal}
  {\bibinfo  {journal} {EPL (Europhysics Letters)}\ }\textbf {\bibinfo {volume}
  {50}},\ \bibinfo {pages} {688} (\bibinfo {year} {2000})}\BibitemShut
  {NoStop}%
\bibitem [{\citenamefont {B{\"a}uerle}\ \emph {et~al.}(1980)\citenamefont
  {B{\"a}uerle}, \citenamefont {Wagner}, \citenamefont {W{\"o}hlecke},
  \citenamefont {Dorner},\ and\ \citenamefont {Kraxenberger}}]{Bauerle1980}%
  \BibitemOpen
  \bibfield  {author} {\bibinfo {author} {\bibfnamefont {D.}~\bibnamefont
  {B{\"a}uerle}}, \bibinfo {author} {\bibfnamefont {D.}~\bibnamefont {Wagner}},
  \bibinfo {author} {\bibfnamefont {M.}~\bibnamefont {W{\"o}hlecke}}, \bibinfo
  {author} {\bibfnamefont {B.}~\bibnamefont {Dorner}},\ and\ \bibinfo {author}
  {\bibfnamefont {H.}~\bibnamefont {Kraxenberger}},\ }\bibfield  {title}
  {\bibinfo {title} {Soft modes in semiconducting srtio3: Ii. the ferroelectric
  mode},\ }\href@noop {} {\bibfield  {journal} {\bibinfo  {journal}
  {Zeitschrift f{\"u}r Physik B Condensed Matter}\ }\textbf {\bibinfo {volume}
  {38}},\ \bibinfo {pages} {335} (\bibinfo {year} {1980})}\BibitemShut
  {NoStop}%
\bibitem [{\citenamefont {Prakash}\ \emph {et~al.}(2017)\citenamefont
  {Prakash}, \citenamefont {Kumar}, \citenamefont {Thamizhavel},\ and\
  \citenamefont {Ramakrishnan}}]{prakash2017evidence}%
  \BibitemOpen
  \bibfield  {author} {\bibinfo {author} {\bibfnamefont {O.}~\bibnamefont
  {Prakash}}, \bibinfo {author} {\bibfnamefont {A.}~\bibnamefont {Kumar}},
  \bibinfo {author} {\bibfnamefont {A.}~\bibnamefont {Thamizhavel}},\ and\
  \bibinfo {author} {\bibfnamefont {S.}~\bibnamefont {Ramakrishnan}},\
  }\bibfield  {title} {\bibinfo {title} {Evidence for bulk superconductivity in
  pure bismuth single crystals at ambient pressure},\ }\href@noop {} {\bibfield
   {journal} {\bibinfo  {journal} {Science}\ }\textbf {\bibinfo {volume}
  {355}},\ \bibinfo {pages} {52} (\bibinfo {year} {2017})}\BibitemShut
  {NoStop}%
\bibitem [{\citenamefont {Bretz-Sullivan}\ \emph {et~al.}(2019)\citenamefont
  {Bretz-Sullivan}, \citenamefont {Edelman}, \citenamefont {Jiang},
  \citenamefont {Suslov}, \citenamefont {Graf}, \citenamefont {Zhang},
  \citenamefont {Wang}, \citenamefont {Chang}, \citenamefont {Pearson},
  \citenamefont {Martinson} \emph {et~al.}}]{bretz2019superconductivity}%
  \BibitemOpen
  \bibfield  {author} {\bibinfo {author} {\bibfnamefont {T.~M.}\ \bibnamefont
  {Bretz-Sullivan}}, \bibinfo {author} {\bibfnamefont {A.}~\bibnamefont
  {Edelman}}, \bibinfo {author} {\bibfnamefont {J.}~\bibnamefont {Jiang}},
  \bibinfo {author} {\bibfnamefont {A.}~\bibnamefont {Suslov}}, \bibinfo
  {author} {\bibfnamefont {D.}~\bibnamefont {Graf}}, \bibinfo {author}
  {\bibfnamefont {J.}~\bibnamefont {Zhang}}, \bibinfo {author} {\bibfnamefont
  {G.}~\bibnamefont {Wang}}, \bibinfo {author} {\bibfnamefont {C.}~\bibnamefont
  {Chang}}, \bibinfo {author} {\bibfnamefont {J.~E.}\ \bibnamefont {Pearson}},
  \bibinfo {author} {\bibfnamefont {A.~B.}\ \bibnamefont {Martinson}}, \emph
  {et~al.},\ }\bibfield  {title} {\bibinfo {title} {Superconductivity in the
  dilute single band limit in reduced strontium titanate},\ }\href@noop {}
  {\bibfield  {journal} {\bibinfo  {journal} {arXiv preprint arXiv:1904.03121}\
  } (\bibinfo {year} {2019})}\BibitemShut {NoStop}%
\bibitem [{\citenamefont {Koonce}\ \emph {et~al.}(1967)\citenamefont {Koonce},
  \citenamefont {Cohen}, \citenamefont {Schooley}, \citenamefont {Hosler},\
  and\ \citenamefont {Pfeiffer}}]{Koonce1967}%
  \BibitemOpen
  \bibfield  {author} {\bibinfo {author} {\bibfnamefont {C.~S.}\ \bibnamefont
  {Koonce}}, \bibinfo {author} {\bibfnamefont {M.~L.}\ \bibnamefont {Cohen}},
  \bibinfo {author} {\bibfnamefont {J.~F.}\ \bibnamefont {Schooley}}, \bibinfo
  {author} {\bibfnamefont {W.~R.}\ \bibnamefont {Hosler}},\ and\ \bibinfo
  {author} {\bibfnamefont {E.~R.}\ \bibnamefont {Pfeiffer}},\ }\bibfield
  {title} {\bibinfo {title} {{Superconducting Transition Temperatures of
  Semiconducting SrTiO3}},\ }\href {https://doi.org/10.1103/PhysRev.163.380}
  {\bibfield  {journal} {\bibinfo  {journal} {Phys. Rev.}\ }\textbf {\bibinfo
  {volume} {163}},\ \bibinfo {pages} {380} (\bibinfo {year}
  {1967})}\BibitemShut {NoStop}%
\bibitem [{\citenamefont {Collignon}\ \emph {et~al.}(2017)\citenamefont
  {Collignon}, \citenamefont {Fauqu{\'e}}, \citenamefont {Cavanna},
  \citenamefont {Gennser}, \citenamefont {Mailly},\ and\ \citenamefont
  {Behnia}}]{Collignon2017}%
  \BibitemOpen
  \bibfield  {author} {\bibinfo {author} {\bibfnamefont {C.}~\bibnamefont
  {Collignon}}, \bibinfo {author} {\bibfnamefont {B.}~\bibnamefont
  {Fauqu{\'e}}}, \bibinfo {author} {\bibfnamefont {A.}~\bibnamefont {Cavanna}},
  \bibinfo {author} {\bibfnamefont {U.}~\bibnamefont {Gennser}}, \bibinfo
  {author} {\bibfnamefont {D.}~\bibnamefont {Mailly}},\ and\ \bibinfo {author}
  {\bibfnamefont {K.}~\bibnamefont {Behnia}},\ }\bibfield  {title} {\bibinfo
  {title} {Superfluid density and carrier concentration across a
  superconducting dome: The case of strontium titanate},\ }\href@noop {}
  {\bibfield  {journal} {\bibinfo  {journal} {Physical Review B}\ }\textbf
  {\bibinfo {volume} {96}},\ \bibinfo {pages} {224506} (\bibinfo {year}
  {2017})}\BibitemShut {NoStop}%
\bibitem [{\citenamefont {Collignon}\ \emph {et~al.}(2019)\citenamefont
  {Collignon}, \citenamefont {Lin}, \citenamefont {Rischau}, \citenamefont
  {Fauqu{\'e}},\ and\ \citenamefont {Behnia}}]{collignon2019}%
  \BibitemOpen
  \bibfield  {author} {\bibinfo {author} {\bibfnamefont {C.}~\bibnamefont
  {Collignon}}, \bibinfo {author} {\bibfnamefont {X.}~\bibnamefont {Lin}},
  \bibinfo {author} {\bibfnamefont {C.~W.}\ \bibnamefont {Rischau}}, \bibinfo
  {author} {\bibfnamefont {B.}~\bibnamefont {Fauqu{\'e}}},\ and\ \bibinfo
  {author} {\bibfnamefont {K.}~\bibnamefont {Behnia}},\ }\bibfield  {title}
  {\bibinfo {title} {Metallicity and superconductivity in doped strontium
  titanate},\ }\href@noop {} {\bibfield  {journal} {\bibinfo  {journal} {Annual
  Review of Condensed Matter Physics}\ }\textbf {\bibinfo {volume} {10}},\
  \bibinfo {pages} {25} (\bibinfo {year} {2019})}\BibitemShut {NoStop}%
\bibitem [{\citenamefont {McCalla}\ \emph {et~al.}(2019)\citenamefont
  {McCalla}, \citenamefont {Gastiasoro}, \citenamefont {Cassuto}, \citenamefont
  {Fernandes},\ and\ \citenamefont {Leighton}}]{McCalla2019}%
  \BibitemOpen
  \bibfield  {author} {\bibinfo {author} {\bibfnamefont {E.}~\bibnamefont
  {McCalla}}, \bibinfo {author} {\bibfnamefont {M.~N.}\ \bibnamefont
  {Gastiasoro}}, \bibinfo {author} {\bibfnamefont {G.}~\bibnamefont {Cassuto}},
  \bibinfo {author} {\bibfnamefont {R.~M.}\ \bibnamefont {Fernandes}},\ and\
  \bibinfo {author} {\bibfnamefont {C.}~\bibnamefont {Leighton}},\ }\bibfield
  {title} {\bibinfo {title} {Low-temperature specific heat of doped
  $\mathrm{SrTi}{\mathrm{o}}_{3}$: Doping dependence of the effective mass and
  kadowaki-woods scaling violation},\ }\href
  {https://doi.org/10.1103/PhysRevMaterials.3.022001} {\bibfield  {journal}
  {\bibinfo  {journal} {Phys. Rev. Materials}\ }\textbf {\bibinfo {volume}
  {3}},\ \bibinfo {pages} {022001} (\bibinfo {year} {2019})}\BibitemShut
  {NoStop}%
\bibitem [{\citenamefont {Lin}\ \emph {et~al.}(2015)\citenamefont {Lin},
  \citenamefont {Rischau}, \citenamefont {van~der Beek}, \citenamefont
  {Fauqu{\'e}},\ and\ \citenamefont {Behnia}}]{lin2015s}%
  \BibitemOpen
  \bibfield  {author} {\bibinfo {author} {\bibfnamefont {X.}~\bibnamefont
  {Lin}}, \bibinfo {author} {\bibfnamefont {C.~W.}\ \bibnamefont {Rischau}},
  \bibinfo {author} {\bibfnamefont {C.~J.}\ \bibnamefont {van~der Beek}},
  \bibinfo {author} {\bibfnamefont {B.}~\bibnamefont {Fauqu{\'e}}},\ and\
  \bibinfo {author} {\bibfnamefont {K.}~\bibnamefont {Behnia}},\ }\bibfield
  {title} {\bibinfo {title} {s-wave superconductivity in optimally doped srti
  1- x nb x o 3 unveiled by electron irradiation},\ }\href@noop {} {\bibfield
  {journal} {\bibinfo  {journal} {Physical Review B}\ }\textbf {\bibinfo
  {volume} {92}},\ \bibinfo {pages} {174504} (\bibinfo {year}
  {2015})}\BibitemShut {NoStop}%
\bibitem [{\citenamefont {Dec}\ \emph {et~al.}(2004)\citenamefont {Dec},
  \citenamefont {Kleemann},\ and\ \citenamefont {Itoh}}]{Dec2004}%
  \BibitemOpen
  \bibfield  {author} {\bibinfo {author} {\bibfnamefont {J.}~\bibnamefont
  {Dec}}, \bibinfo {author} {\bibfnamefont {W.}~\bibnamefont {Kleemann}},\ and\
  \bibinfo {author} {\bibfnamefont {M.}~\bibnamefont {Itoh}},\ }\bibfield
  {title} {\bibinfo {title} {Electric-field-induced ferroelastic single
  domaining of sr ti o 3 18},\ }\href@noop {} {\bibfield  {journal} {\bibinfo
  {journal} {Applied physics letters}\ }\textbf {\bibinfo {volume} {85}},\
  \bibinfo {pages} {5328} (\bibinfo {year} {2004})}\BibitemShut {NoStop}%
\bibitem [{\citenamefont {Caprara}\ \emph {et~al.}(2020)\citenamefont
  {Caprara}, \citenamefont {Grilli}, \citenamefont {Lorenzana},\ and\
  \citenamefont {Leridon}}]{Caprara2020}%
  \BibitemOpen
  \bibfield  {author} {\bibinfo {author} {\bibfnamefont {S.}~\bibnamefont
  {Caprara}}, \bibinfo {author} {\bibfnamefont {M.}~\bibnamefont {Grilli}},
  \bibinfo {author} {\bibfnamefont {J.}~\bibnamefont {Lorenzana}},\ and\
  \bibinfo {author} {\bibfnamefont {B.}~\bibnamefont {Leridon}},\ }\bibfield
  {title} {\bibinfo {title} {Doping-dependent competition between
  superconductivity and polycrystalline charge density waves},\ }\href@noop {}
  {\bibfield  {journal} {\bibinfo  {journal} {SciPost Physics}\ }\textbf
  {\bibinfo {volume} {8}},\ \bibinfo {pages} {003} (\bibinfo {year}
  {2020})}\BibitemShut {NoStop}%
\bibitem [{\citenamefont {Leridon}\ \emph {et~al.}(2020)\citenamefont
  {Leridon}, \citenamefont {Caprara}, \citenamefont {Vanacken}, \citenamefont
  {Moshchalkov}, \citenamefont {Vignolle}, \citenamefont {Porwal},
  \citenamefont {Budhani}, \citenamefont {Attanasi}, \citenamefont {Grilli},\
  and\ \citenamefont {Lorenzana}}]{Leridon2020}%
  \BibitemOpen
  \bibfield  {author} {\bibinfo {author} {\bibfnamefont {B.}~\bibnamefont
  {Leridon}}, \bibinfo {author} {\bibfnamefont {S.}~\bibnamefont {Caprara}},
  \bibinfo {author} {\bibfnamefont {J.}~\bibnamefont {Vanacken}}, \bibinfo
  {author} {\bibfnamefont {V.}~\bibnamefont {Moshchalkov}}, \bibinfo {author}
  {\bibfnamefont {B.}~\bibnamefont {Vignolle}}, \bibinfo {author}
  {\bibfnamefont {R.}~\bibnamefont {Porwal}}, \bibinfo {author} {\bibfnamefont
  {R.}~\bibnamefont {Budhani}}, \bibinfo {author} {\bibfnamefont
  {A.}~\bibnamefont {Attanasi}}, \bibinfo {author} {\bibfnamefont
  {M.}~\bibnamefont {Grilli}},\ and\ \bibinfo {author} {\bibfnamefont
  {J.}~\bibnamefont {Lorenzana}},\ }\bibfield  {title} {\bibinfo {title}
  {Protected superconductivity at the boundaries of charge-density-wave
  domains},\ }\href@noop {} {\bibfield  {journal} {\bibinfo  {journal} {New
  Journal of Physics}\ }\textbf {\bibinfo {volume} {22}},\ \bibinfo {pages}
  {073025} (\bibinfo {year} {2020})}\BibitemShut {NoStop}%
\bibitem [{\citenamefont {Szot}\ \emph {et~al.}(2021)\citenamefont {Szot},
  \citenamefont {Rodenb{\"u}cher}, \citenamefont {Rogacki}, \citenamefont
  {Bihlmayer}, \citenamefont {Speier}, \citenamefont {Roleder}, \citenamefont
  {Krok}, \citenamefont {Keller}, \citenamefont {Simon},\ and\ \citenamefont
  {Bussmann-Holder}}]{szot2021filamentary}%
  \BibitemOpen
  \bibfield  {author} {\bibinfo {author} {\bibfnamefont {K.}~\bibnamefont
  {Szot}}, \bibinfo {author} {\bibfnamefont {C.}~\bibnamefont
  {Rodenb{\"u}cher}}, \bibinfo {author} {\bibfnamefont {K.}~\bibnamefont
  {Rogacki}}, \bibinfo {author} {\bibfnamefont {G.}~\bibnamefont {Bihlmayer}},
  \bibinfo {author} {\bibfnamefont {W.}~\bibnamefont {Speier}}, \bibinfo
  {author} {\bibfnamefont {K.}~\bibnamefont {Roleder}}, \bibinfo {author}
  {\bibfnamefont {F.}~\bibnamefont {Krok}}, \bibinfo {author} {\bibfnamefont
  {H.}~\bibnamefont {Keller}}, \bibinfo {author} {\bibfnamefont
  {A.}~\bibnamefont {Simon}},\ and\ \bibinfo {author} {\bibfnamefont
  {A.}~\bibnamefont {Bussmann-Holder}},\ }\bibfield  {title} {\bibinfo {title}
  {Filamentary superconductivity of resistively-switched strontium titanate},\
  }\href@noop {} {\bibfield  {journal} {\bibinfo  {journal} {arXiv preprint
  arXiv:2110.07230}\ } (\bibinfo {year} {2021})}\BibitemShut {NoStop}%
\bibitem [{\citenamefont {Caprara}\ \emph {et~al.}(2013)\citenamefont
  {Caprara}, \citenamefont {Biscaras}, \citenamefont {Bergeal}, \citenamefont
  {Bucheli}, \citenamefont {Hurand}, \citenamefont {Feuillet-Palma},
  \citenamefont {Rastogi}, \citenamefont {Budhani}, \citenamefont {Lesueur},\
  and\ \citenamefont {Grilli}}]{Caprara2013}%
  \BibitemOpen
  \bibfield  {author} {\bibinfo {author} {\bibfnamefont {S.}~\bibnamefont
  {Caprara}}, \bibinfo {author} {\bibfnamefont {J.}~\bibnamefont {Biscaras}},
  \bibinfo {author} {\bibfnamefont {N.}~\bibnamefont {Bergeal}}, \bibinfo
  {author} {\bibfnamefont {D.}~\bibnamefont {Bucheli}}, \bibinfo {author}
  {\bibfnamefont {S.}~\bibnamefont {Hurand}}, \bibinfo {author} {\bibfnamefont
  {C.}~\bibnamefont {Feuillet-Palma}}, \bibinfo {author} {\bibfnamefont
  {A.}~\bibnamefont {Rastogi}}, \bibinfo {author} {\bibfnamefont {R.~C.}\
  \bibnamefont {Budhani}}, \bibinfo {author} {\bibfnamefont {J.}~\bibnamefont
  {Lesueur}},\ and\ \bibinfo {author} {\bibfnamefont {M.}~\bibnamefont
  {Grilli}},\ }\bibfield  {title} {\bibinfo {title} {{Multiband
  superconductivity and nanoscale inhomogeneity at oxide interfaces}},\ }\href
  {https://doi.org/10.1103/PhysRevB.88.020504} {\bibfield  {journal} {\bibinfo
  {journal} {Phys. Rev. B}\ }\textbf {\bibinfo {volume} {88}},\ \bibinfo
  {pages} {020504} (\bibinfo {year} {2013})}\BibitemShut {NoStop}%
\bibitem [{\citenamefont {Reyren}\ \emph {et~al.}(2007)\citenamefont {Reyren},
  \citenamefont {Thiel}, \citenamefont {Caviglia}, \citenamefont {Kourkoutis},
  \citenamefont {Hammerl}, \citenamefont {Richter}, \citenamefont {Schneider},
  \citenamefont {Kopp}, \citenamefont {R{\"u}etschi}, \citenamefont {Jaccard}
  \emph {et~al.}}]{reyren2007}%
  \BibitemOpen
  \bibfield  {author} {\bibinfo {author} {\bibfnamefont {N.}~\bibnamefont
  {Reyren}}, \bibinfo {author} {\bibfnamefont {S.}~\bibnamefont {Thiel}},
  \bibinfo {author} {\bibfnamefont {A.}~\bibnamefont {Caviglia}}, \bibinfo
  {author} {\bibfnamefont {L.~F.}\ \bibnamefont {Kourkoutis}}, \bibinfo
  {author} {\bibfnamefont {G.}~\bibnamefont {Hammerl}}, \bibinfo {author}
  {\bibfnamefont {C.}~\bibnamefont {Richter}}, \bibinfo {author} {\bibfnamefont
  {C.~W.}\ \bibnamefont {Schneider}}, \bibinfo {author} {\bibfnamefont
  {T.}~\bibnamefont {Kopp}}, \bibinfo {author} {\bibfnamefont {A.-S.}\
  \bibnamefont {R{\"u}etschi}}, \bibinfo {author} {\bibfnamefont
  {D.}~\bibnamefont {Jaccard}}, \emph {et~al.},\ }\bibfield  {title} {\bibinfo
  {title} {Superconducting interfaces between insulating oxides},\ }\href@noop
  {} {\bibfield  {journal} {\bibinfo  {journal} {Science}\ }\textbf {\bibinfo
  {volume} {317}},\ \bibinfo {pages} {1196} (\bibinfo {year}
  {2007})}\BibitemShut {NoStop}%
\bibitem [{\citenamefont {Han}\ \emph {et~al.}(2014)\citenamefont {Han},
  \citenamefont {Shen}, \citenamefont {You}, \citenamefont {Li}, \citenamefont
  {Luo}, \citenamefont {Li}, \citenamefont {Qu}, \citenamefont {Xiong},
  \citenamefont {Dou}, \citenamefont {He} \emph {et~al.}}]{han2014two}%
  \BibitemOpen
  \bibfield  {author} {\bibinfo {author} {\bibfnamefont {Y.-L.}\ \bibnamefont
  {Han}}, \bibinfo {author} {\bibfnamefont {S.-C.}\ \bibnamefont {Shen}},
  \bibinfo {author} {\bibfnamefont {J.}~\bibnamefont {You}}, \bibinfo {author}
  {\bibfnamefont {H.-O.}\ \bibnamefont {Li}}, \bibinfo {author} {\bibfnamefont
  {Z.-Z.}\ \bibnamefont {Luo}}, \bibinfo {author} {\bibfnamefont {C.-J.}\
  \bibnamefont {Li}}, \bibinfo {author} {\bibfnamefont {G.-L.}\ \bibnamefont
  {Qu}}, \bibinfo {author} {\bibfnamefont {C.-M.}\ \bibnamefont {Xiong}},
  \bibinfo {author} {\bibfnamefont {R.-F.}\ \bibnamefont {Dou}}, \bibinfo
  {author} {\bibfnamefont {L.}~\bibnamefont {He}}, \emph {et~al.},\ }\bibfield
  {title} {\bibinfo {title} {Two-dimensional superconductivity at (110)
  laalo3/srtio3 interfaces},\ }\href@noop {} {\bibfield  {journal} {\bibinfo
  {journal} {Applied Physics Letters}\ }\textbf {\bibinfo {volume} {105}},\
  \bibinfo {pages} {192603} (\bibinfo {year} {2014})}\BibitemShut {NoStop}%
\bibitem [{\citenamefont {Liu}\ \emph {et~al.}(2021)\citenamefont {Liu},
  \citenamefont {Yan}, \citenamefont {Jin}, \citenamefont {Ma}, \citenamefont
  {Hsiao}, \citenamefont {Lin}, \citenamefont {Bretz-Sullivan}, \citenamefont
  {Zhou}, \citenamefont {Pearson}, \citenamefont {Fisher} \emph
  {et~al.}}]{liu2021two}%
  \BibitemOpen
  \bibfield  {author} {\bibinfo {author} {\bibfnamefont {C.}~\bibnamefont
  {Liu}}, \bibinfo {author} {\bibfnamefont {X.}~\bibnamefont {Yan}}, \bibinfo
  {author} {\bibfnamefont {D.}~\bibnamefont {Jin}}, \bibinfo {author}
  {\bibfnamefont {Y.}~\bibnamefont {Ma}}, \bibinfo {author} {\bibfnamefont
  {H.-W.}\ \bibnamefont {Hsiao}}, \bibinfo {author} {\bibfnamefont
  {Y.}~\bibnamefont {Lin}}, \bibinfo {author} {\bibfnamefont {T.~M.}\
  \bibnamefont {Bretz-Sullivan}}, \bibinfo {author} {\bibfnamefont
  {X.}~\bibnamefont {Zhou}}, \bibinfo {author} {\bibfnamefont {J.}~\bibnamefont
  {Pearson}}, \bibinfo {author} {\bibfnamefont {B.}~\bibnamefont {Fisher}},
  \emph {et~al.},\ }\bibfield  {title} {\bibinfo {title} {Two-dimensional
  superconductivity and anisotropic transport at ktao3 (111) interfaces},\
  }\href@noop {} {\bibfield  {journal} {\bibinfo  {journal} {Science}\ }\textbf
  {\bibinfo {volume} {371}},\ \bibinfo {pages} {716} (\bibinfo {year}
  {2021})}\BibitemShut {NoStop}%
\bibitem [{\citenamefont {Axe}(1967)}]{1967Axe}%
  \BibitemOpen
  \bibfield  {author} {\bibinfo {author} {\bibfnamefont {J.~D.}\ \bibnamefont
  {Axe}},\ }\bibfield  {title} {\bibinfo {title} {Apparent ionic charges and
  vibrational eigenmodes of bati${\mathrm{o}}_{3}$ and other perovskites},\
  }\href {https://doi.org/10.1103/PhysRev.157.429} {\bibfield  {journal}
  {\bibinfo  {journal} {Phys. Rev.}\ }\textbf {\bibinfo {volume} {157}},\
  \bibinfo {pages} {429} (\bibinfo {year} {1967})}\BibitemShut {NoStop}%
\bibitem [{\citenamefont {Kresse}\ and\ \citenamefont
  {Furthm{\"u}ller}(1996)}]{vasp1}%
  \BibitemOpen
  \bibfield  {author} {\bibinfo {author} {\bibfnamefont {G.}~\bibnamefont
  {Kresse}}\ and\ \bibinfo {author} {\bibfnamefont {J.}~\bibnamefont
  {Furthm{\"u}ller}},\ }\bibfield  {title} {\bibinfo {title} {Efficient
  iterative schemes for ab initio total-energy calculations using a plane-wave
  basis set},\ }\href@noop {} {\bibfield  {journal} {\bibinfo  {journal}
  {Physical review B}\ }\textbf {\bibinfo {volume} {54}},\ \bibinfo {pages}
  {11169} (\bibinfo {year} {1996})}\BibitemShut {NoStop}%
\bibitem [{\citenamefont {Kresse}\ and\ \citenamefont {Joubert}(1999)}]{vasp2}%
  \BibitemOpen
  \bibfield  {author} {\bibinfo {author} {\bibfnamefont {G.}~\bibnamefont
  {Kresse}}\ and\ \bibinfo {author} {\bibfnamefont {D.}~\bibnamefont
  {Joubert}},\ }\bibfield  {title} {\bibinfo {title} {From ultrasoft
  pseudopotentials to the projector augmented-wave method},\ }\href@noop {}
  {\bibfield  {journal} {\bibinfo  {journal} {Physical review b}\ }\textbf
  {\bibinfo {volume} {59}},\ \bibinfo {pages} {1758} (\bibinfo {year}
  {1999})}\BibitemShut {NoStop}%
\bibitem [{\citenamefont {Perdew}\ \emph {et~al.}(2008)\citenamefont {Perdew},
  \citenamefont {Ruzsinszky}, \citenamefont {Csonka}, \citenamefont {Vydrov},
  \citenamefont {Scuseria}, \citenamefont {Constantin}, \citenamefont {Zhou},\
  and\ \citenamefont {Burke}}]{pbesol}%
  \BibitemOpen
  \bibfield  {author} {\bibinfo {author} {\bibfnamefont {J.~P.}\ \bibnamefont
  {Perdew}}, \bibinfo {author} {\bibfnamefont {A.}~\bibnamefont {Ruzsinszky}},
  \bibinfo {author} {\bibfnamefont {G.~I.}\ \bibnamefont {Csonka}}, \bibinfo
  {author} {\bibfnamefont {O.~A.}\ \bibnamefont {Vydrov}}, \bibinfo {author}
  {\bibfnamefont {G.~E.}\ \bibnamefont {Scuseria}}, \bibinfo {author}
  {\bibfnamefont {L.~A.}\ \bibnamefont {Constantin}}, \bibinfo {author}
  {\bibfnamefont {X.}~\bibnamefont {Zhou}},\ and\ \bibinfo {author}
  {\bibfnamefont {K.}~\bibnamefont {Burke}},\ }\bibfield  {title} {\bibinfo
  {title} {Restoring the density-gradient expansion for exchange in solids and
  surfaces},\ }\href@noop {} {\bibfield  {journal} {\bibinfo  {journal}
  {Physical review letters}\ }\textbf {\bibinfo {volume} {100}},\ \bibinfo
  {pages} {136406} (\bibinfo {year} {2008})}\BibitemShut {NoStop}%
\bibitem [{\citenamefont {Lytle}(1964)}]{xray}%
  \BibitemOpen
  \bibfield  {author} {\bibinfo {author} {\bibfnamefont {F.~W.}\ \bibnamefont
  {Lytle}},\ }\bibfield  {title} {\bibinfo {title} {X-ray diffractometry of
  low-temperature phase transformations in strontium titanate},\ }\href@noop {}
  {\bibfield  {journal} {\bibinfo  {journal} {Journal of Applied Physics}\
  }\textbf {\bibinfo {volume} {35}},\ \bibinfo {pages} {2212} (\bibinfo {year}
  {1964})}\BibitemShut {NoStop}%
\bibitem [{\citenamefont {van~der Marel}\ \emph {et~al.}(2011)\citenamefont
  {van~der Marel}, \citenamefont {van Mechelen},\ and\ \citenamefont
  {Mazin}}]{vanderMarel2011}%
  \BibitemOpen
  \bibfield  {author} {\bibinfo {author} {\bibfnamefont {D.}~\bibnamefont
  {van~der Marel}}, \bibinfo {author} {\bibfnamefont {J.~L.~M.}\ \bibnamefont
  {van Mechelen}},\ and\ \bibinfo {author} {\bibfnamefont {I.~I.}\ \bibnamefont
  {Mazin}},\ }\bibfield  {title} {\bibinfo {title} {Common fermi-liquid origin
  of ${T}^{2}$ resistivity and superconductivity in $n$-type srtio${}_{3}$},\
  }\href {https://doi.org/10.1103/PhysRevB.84.205111} {\bibfield  {journal}
  {\bibinfo  {journal} {Phys. Rev. B}\ }\textbf {\bibinfo {volume} {84}},\
  \bibinfo {pages} {205111} (\bibinfo {year} {2011})}\BibitemShut {NoStop}%
\bibitem [{\citenamefont {Gao}\ \emph {et~al.}(2018)\citenamefont {Gao},
  \citenamefont {Yang}, \citenamefont {Ishikawa}, \citenamefont {Li},
  \citenamefont {Feng}, \citenamefont {Kumamoto}, \citenamefont {Shibata},
  \citenamefont {Yu},\ and\ \citenamefont {Ikuhara}}]{gao2018atomic}%
  \BibitemOpen
  \bibfield  {author} {\bibinfo {author} {\bibfnamefont {P.}~\bibnamefont
  {Gao}}, \bibinfo {author} {\bibfnamefont {S.}~\bibnamefont {Yang}}, \bibinfo
  {author} {\bibfnamefont {R.}~\bibnamefont {Ishikawa}}, \bibinfo {author}
  {\bibfnamefont {N.}~\bibnamefont {Li}}, \bibinfo {author} {\bibfnamefont
  {B.}~\bibnamefont {Feng}}, \bibinfo {author} {\bibfnamefont {A.}~\bibnamefont
  {Kumamoto}}, \bibinfo {author} {\bibfnamefont {N.}~\bibnamefont {Shibata}},
  \bibinfo {author} {\bibfnamefont {P.}~\bibnamefont {Yu}},\ and\ \bibinfo
  {author} {\bibfnamefont {Y.}~\bibnamefont {Ikuhara}},\ }\bibfield  {title}
  {\bibinfo {title} {Atomic-scale measurement of flexoelectric polarization at
  srtio 3 dislocations},\ }\href@noop {} {\bibfield  {journal} {\bibinfo
  {journal} {Physical review letters}\ }\textbf {\bibinfo {volume} {120}},\
  \bibinfo {pages} {267601} (\bibinfo {year} {2018})}\BibitemShut {NoStop}%
\bibitem [{\citenamefont {M{\"u}ller}\ and\ \citenamefont
  {Wyart}(2015)}]{muller2015marginal}%
  \BibitemOpen
  \bibfield  {author} {\bibinfo {author} {\bibfnamefont {M.}~\bibnamefont
  {M{\"u}ller}}\ and\ \bibinfo {author} {\bibfnamefont {M.}~\bibnamefont
  {Wyart}},\ }\bibfield  {title} {\bibinfo {title} {Marginal stability in
  structural, spin, and electron glasses},\ }\href@noop {} {\bibfield
  {journal} {\bibinfo  {journal} {Annu. Rev. Condens. Matter Phys.}\ }\textbf
  {\bibinfo {volume} {6}},\ \bibinfo {pages} {177} (\bibinfo {year}
  {2015})}\BibitemShut {NoStop}%
\bibitem [{\citenamefont {Franz}\ \emph {et~al.}(2015)\citenamefont {Franz},
  \citenamefont {Parisi}, \citenamefont {Urbani},\ and\ \citenamefont
  {Zamponi}}]{franz2015universal}%
  \BibitemOpen
  \bibfield  {author} {\bibinfo {author} {\bibfnamefont {S.}~\bibnamefont
  {Franz}}, \bibinfo {author} {\bibfnamefont {G.}~\bibnamefont {Parisi}},
  \bibinfo {author} {\bibfnamefont {P.}~\bibnamefont {Urbani}},\ and\ \bibinfo
  {author} {\bibfnamefont {F.}~\bibnamefont {Zamponi}},\ }\bibfield  {title}
  {\bibinfo {title} {Universal spectrum of normal modes in low-temperature
  glasses},\ }\href@noop {} {\bibfield  {journal} {\bibinfo  {journal}
  {Proceedings of the National Academy of Sciences}\ }\textbf {\bibinfo
  {volume} {112}},\ \bibinfo {pages} {14539} (\bibinfo {year}
  {2015})}\BibitemShut {NoStop}%
\end{thebibliography}%

\end{document}